\documentclass[12pt]{article}
\usepackage{comment}
\usepackage{color}
\usepackage{graphicx}
\usepackage{epstopdf}
\usepackage{amsfonts}
\usepackage{amssymb}
\usepackage{amsmath}
\usepackage{harvard}
\usepackage{rotating}
\usepackage{array,booktabs,tabularx}
\newcolumntype{C}{>{\centering\arraybackslash}X}

\newtheorem{pr}{Proposition}
\newtheorem{lem}{Lemma}[section]

\newtheorem{exa}{Example}[section]

\newtheorem{cor}{Corollary}[section]
\newtheorem{df}{Definition}

\def\P{{\cal P}}

\def\be{\begin{equation}} 
\def\ee{\end{equation}} 
\def\beqn{\begin{eqnarray}} 
\def\eeqn{\end{eqnarray}} 
\def\beq{\begin{eqnarray*}} 
\def\eeq{\end{eqnarray*}} 
\def\ba{\begin{array}} 
\def\ea{\end{array}} 
\newcommand{\bex}{\begin{exa}}
\newcommand{\eex}{\end{exa}\vspace{-4mm}}

\newcommand{\br}{\begin{re}}
\newcommand{\er}{\end{re}\vspace{-4mm}}

\definecolor{myBlue}{rgb}{0.80,0.85,1.00}
\definecolor{myYellow}{rgb}{0,1.000,0}

\newbox\treebox
\def\tree{\global\setbox\treebox=\boxtree}

\def\endsubtree{\ettext \egroup}

\newif\iftreetext\treetextfalse
\def\boxtree{\hbox\bgroup
  \baselineskip 2.5ex
  \tabskip 0pt
  \vbox\bgroup
  \treetexttrue
  \let\par\crcr \obeylines
  \halign\bgroup##\hfil\cr}
\def\ettext{\iftreetext
  \crcr\egroup \egroup \fi}

\def\cons#1#2{\edef#2{\xmark #1#2}}
\def\car#1{\expandafter\docar#1\docar}
\def\docar\xmark#1\xmark#2\docar{#1}
\def\cdr#1{\expandafter\docdr#1\docdr#1}
\def\docdr\xmark#1\xmark#2\docdr#3{\def#3{\xmark #2}}
\def\xmark{\noexpand\xmark}
\def\nil{\xmark}

\def\settreesizes{\setbox0=\copy\treebox \global\let\treesizes\nil \setsizes}
\newdimen\treewidth
\def\setsizes{\setbox0=\hbox\bgroup
  \unhbox0\unskip
  \inittreewidth
  \sizesubtrees
  \sizelevel
  \egroup}

\def\inittreewidth{\ifx\treesizes\nil   
  \treewidth=0pt                        
\else \treewidth=\car\treesizes         
  \global\cdr\treesizes                 
  \fi}                                  

\def\sizesubtrees{\loop                 
  \setbox0=\lastbox \unskip             
  \ifhbox0 \setsizes                    
  \repeat}                              

\def\sizelevel{\ifdim\treewidth<\wd0    
  \treewidth=\wd0 \fi                   
\global\cons{\the\treewidth}\treesizes} 

\newdimen\treeheight                    
\newif\ifleaf                           
\newif\ifbotsub                         
\newif\iftopsub                         
\def\maketree{\hbox{\treewidth=\car\treesizes  
  \cdr\treesizes                        
  \makesubtreebox\unskip                
  \ifleaf \makeleaf                     
  \else \makeparent \fi}}               

{\catcode`@=11                          
\gdef\makesubtreebox{\unhbox\treebox    
  \unskip\global\setbox\treebox\lastbox 
  \ifvbox\treebox                       
    \global\leaftrue \let\next\relax    
  \else \botsubtrue                     
    \setbox0\box\voidb@x                
    \botsubtrue \let\next\makesubtree   
  \fi \next}}                           

\def\makesubtree{\setbox1\maketree      
  \unskip\global\setbox\treebox\lastbox 
  \treeheight=\ht1                      
  \advance\treeheight 2ex               
  \ifhbox\treebox \topsubfalse          
    \else \topsubtrue \fi               
  \addsubtreebox                        
  \iftopsub \global\leaffalse           
    \let\next\relax \else               
    \botsubfalse \let\next\makesubtree  
  \fi \next}                            

\def\addsubtreebox{\setbox0=\vbox{\subtreebox\unvbox0}}
\def\subtreebox{\hbox\bgroup            
  \vbox to \treeheight\bgroup           
   \ifbotsub \iftopsub \vfil            
       \hrule width 0.4pt               
     \else \treehalfrule \fi \vfil      
   \else \iftopsub \vfil \treehalfrule  
     \else \hrule width 0.4pt height \treeheight \fi\fi 
   \egroup                              
  \treectrbox{\hrule width 1em}\hskip 0.2em\treectrbox{\box1}\egroup}

\def\treectrbox#1{\vbox to \treeheight{\vfil #1\vfil}}
\def\treehalfrule{\dimen0=\treeheight   
  \divide\dimen0 2\advance\dimen0 0.2pt 
  \hrule width 0.4pt height \dimen0}    

\def\makeleaf{\box\treebox}             
\def\makeparent{\ifdim\ht\treebox>\ht0  
  \treeheight=\ht\treebox               
\else \treeheight=\ht0 \fi              
\advance\treewidth-\wd\treebox          
\advance\treewidth 1em                  
\treectrbox{\box\treebox}\hskip 0.2em   
\treectrbox{\hrule width \treewidth}\treectrbox{\box0}} 

\newcommand{\TITLE}
{ Long-Run Law and Entropy
}

\begin{document}

 \pagestyle{plain}

 \nocite{*}
 \bibliographystyle{econometrica}

\pagestyle{plain}

\nocite{*}
\bibliographystyle{econometrica}

\title{\bfseries \TITLE \\
}

\author {
	Weidong Tian\\ University of North Carolina at Charlotte}

\date{}
\renewcommand\footnotemark{}
\title{\Large \bf Long Run Law and Entropy}
\thanks{
	Weidong Tian, Belk College of Business, University of North Carolina at Charlotte. Email addresses: wtian1@uncc.edu.}
\maketitle
%
%
%
%
%
%

\begin{abstract}

		This paper demonstrates the additive and multiplicative version of a long-run law of unexpected shocks for any economic variable. We derive these long-run laws by the martingale theory without relying on the stationary and ergodic conditions. We apply these long-run laws to asset return, risk-adjusted asset return, and the pricing kernel process and derive new asset pricing implications. Moreover, we introduce several dynamic long-term measures on the pricing kernel process, which relies on the sample data of asset return. Finally, we use these long-term measures to diagnose leading asset pricing models. 
		\end{abstract}
		\vspace{1cm}

		

\newpage

\section{Introduction}
\label{sec:introduction}

One of the central assumptions in many leading economics and finance theories is the stationary and ergodic condition for the underlying economic variable(s). This paper presents long-run (asymptotic) properties for a general economic variable by the martingale theory without relying on the stationary and ergodic conditions.\footnote{Despite the great success of ergodic conditions in literature, several nonergodic and nonstationary models have derived important  implications to economics and finance. See, for instance, Durlauf (1993) investigates the nonergodic economy, and Weitzman (2007) studies asset pricing implications in a nonstationary model.}  We develop a long-run theory of unexpected shocks and derive its novel implications from a long-term perspective. 

Specifically, given an economic variable $Y$ that is represented by a process $(Y_t)$, we investigate the following two processes,
\beq
U_{n}(Y) = \frac{\sum_{t=1}^{n} (Y_t - \mathbb{E}_{t-1}[Y_t])}{n}, n \ge 1,
\eeq
and 
\beq
V_{n}(Y) = \prod_{t=1}^{n} \left( \frac{Y_t}{\mathbb{E}_{t-1}[Y_t]} \right)^{\frac{1}{n}}, n \ge 1.
\eeq
Here $Y_t - \mathbb{E}_{t-1}[Y_t]$ is the unexpected shocks between time $t-1$ and $t$, $U_n(Y)$ is the arithmetic average of unexpected shocks of a variable $Y$, and $V_n(Y)$ is its multiplicative variation (geometric average). In the additive long-run law, we show that $U_n(Y) $ converges to zero under a condition that ``{\em the unconditional variances of the unexpected shocks are bounded from above by a finite positive number}". We also show that this condition is both sufficient and necessary to derive meaningful long-run property to relax the stationary and ergodic condition.

This long-run law about $U_n(Y)$ is motivated by the additive Doob-Meyer decomposition of a general stochastic process $(Y_t)$ as follows,
\beq
Y_n = \left\{ \sum_{t=1}^{n} (Y_t - \mathbb{E}_{t-1}[Y_t]) \right\}+ \left\{ \sum_{t=1}^{n} (\mathbb{E}_{t-1}[Y_t] - Y_{t-1}) + Y_0 \right\}.
\eeq
Since the arithmetic average of the martingale (the first term) component in this decomposition converges to zero, the {\em predictable component} (the second term) describes the long-run property of the process $(Y_t)$.

This long-run law is different from the long-run (additive) theory in  Beveridge and Nelson (1981), Hansen and Scheinkman (2009), and Hansen (2012) in several aspects. First, the construction of a {\em permanent (martingale)} component in previous literature relies on specific technical conditions such as underlying ergodic factor even though the variable $Y$ is not. Second, the long-run law in $U_n(Y)$ is about the {\em conditional} expectation and forecasting, while previous literature is mainly about the unconditional element.\footnote{The long-term theory in Hansen and Scheinkman (2009), Hansen (2012) is mostly developed in a continuous-time setting. By its nature the environment is dual to the local theory. In contrast, this paper focuses on the discrete-time framework.} Third, the long-run law offers concrete {\em convergence rate}, and finally, the long-run law implies new long-term measures.

Similarly, $V_n(Y)$ is derived from the following multiplicative Doob-Meyer decomposition,
\beq
Y_n = \left\{  \prod_{t=1}^{n} \frac{Y_t}{\mathbb{E}_{t-1}[Y_t]} \right\} \left\{ Y_0  \prod_{t=1}^{n} \frac{\mathbb{E}_{t-1}[Y_t]}{ Y_{t-1}}\right\}.
\eeq
The paper shows that, under rather weak condition and when $n$ goes to infinity, $V_{n}(Y)$ converges to $e^{-z_{\infty}(Y)}$, where the long-term entropy $z_{\infty}(Y)$ is defined by,
\beq
z_{\infty}(Y) = \lim_{n \rightarrow \infty} \frac{ J_0(Y_1) + J_1(Y_2)  + \cdots + J_{n-1}(Y_n)}{n}, P-a.s,
\eeq
where $
J_t(Y_{t+1}) = log \mathbb{E}_{t}[Y_{t+1}] - \mathbb{E}_{t}[log(Y_{t+1})]$ is the conditional entropy at time $t$. Compared with the long-run law for $U_n(Y)$, the long-term entropy is a {\em convex adjustment} in the long-run law of $V_n(Y)$. 

This multiplicative Doob-Meyer decomposition is closely related to but different from the martingale decomposition studied in Alvarez and Jermann (2005), Hansen (2012), Hansen and Scheinkman (2009), and  Christensen (2017). In these previous studies, the permanent component plays a crucial role, but the temporary component is related to some long-term and unpredictable factors. In contrast, in the long-run law of $V_n(Y)$, the long-term growth rate of the martingale component is characterized by the long-term entropy, and the remaining component is {\em predictable}, which is another insightful factor. 
For instance, for a pricing kernel process, the predictable component is essentially the long forward rate studied in Dybvig, Ingersoll, and Ross (1996). In this regard, we obtain the long-term entropy and the long forward rate in the multiplicative Doob-Meyer decomposition framework. 

Martin (2012) studies the valuation of long-dated asset (for the risk-adjusted asset return $Y$) because the risk-adjusted asset return is a multiplicative martingale. The long-run law of $V_n(Y)$ is also motivated by Martin (2012). This paper derives a refined version of the long-dated asset valuation theory (See implications below).\footnote{This analysis is also related to the tail event analysis in a long term. See Weitzman (2009), and Nordhaus (2011).}
%
%
%
%
%
%

These long-run laws of unexpected shocks are the theoretical building blocks in this paper. 
The class of economic variables considered here is significantly broad, including, for instance, macroeconomic, mortality rate, climate change, weather data, or microeconomic data. Following Hansen (2012) in this paper, we concentrate on the asset return, pricing kernel, and risk-adjusted asset return in applying these long-run laws. We use the apparatus for obtaining new asset pricing implications from a long-term perspective as follows.

{\em First}, for an asset return $R = (R_t)$, we show that the long-term sample excess return equals the long-term expected return, regardless of model assumptions on asset return and subjective probability. Therefore, even though the subjective and objective probability varies, the long-term expected return under any probability measure is the same in the long run. Then, with sample data of index and index options, we investigate whether Martin's (2017) negative correlation condition (NCC) is solid or not. Moreover, the paper demonstrates that the risk-neutral and pessimistic hypothesis (Adam, Matveev, and Nagel (2021)) is theoretically rejected by using a positive equity risk premium.

{\em Second}, for the equity market index return, we demonstrate a close relationship between the long-run law and the no-arbitrage asset pricing theory. We show that the financial crisis event is not a ``{\em Black Swan}" but a natural outcome of the no-arbitrage assumption of the equity market in the long run. Therefore, the long-run law explains the pervasive equity market turmoil phenomenon. Put differently, a long-lived investor in the equity market sees either arbitrage opportunities or persistent market crises.

{\em Third}, for the stochastic discount factor process $m = (m_t)$, the long-run law motivates a long-term measure,
 \beq
\pi(m;s) =\lim_{n \rightarrow \infty} \frac{ \sum_{t=1}^{n} \mathbb{E}_{t-1}[ m_{t}^s ]^{ \frac{1}{1-s}} }{n}, s \in (-\infty, \infty), s \ne 0, s \ne 1.
\eeq
This paper shows that this dynamic measure is bounded (from above or below) by {\em sample} of asset returns. In contrast with widely studied one-period (or conditional) measure of the stochastic discount factor in literature (see e.g. Hansen and Jagannathan (1991), Snow (1991), Bansal and Lehmann (1997), Alvarez and Jermann (2005), and Liu (2021)), $\pi(m;s)$ is defined for the entire stochastic discount process. 

{\em Fourth}, for the risk-adjusted asset return process, the long-run law implies that $(m_1 R_1 \cdots m_n R_n)^{\frac{1}{n}}$ converges to $e^{-z_{\infty}(mR)}$, whereas Martin (2012) shows that $m_1 R_1 \cdots m_n R_n$ converges to zero in a generic sense. Moreover, we show that the Casero sum, $\frac{m_1 R_1 + \cdots + m_nR_n}{n}$, converges to one almost surely, even though $m_nR_n$ diverges in general.

{\em Fifth}, the paper characterizes the long-term entropy of the pricing kernel process in terms of long-term sample excess return (continuously compounding). Therefore, the long-term entropy $z_{\infty}(m)$ is {\em independent} of the specification of the pricing kernel process; instead, it depends only on the {\em sample excess return} of assets. Moreover, we demonstrate the relationship between the long-term entropy with other established long-term measures such as in Hansen (2012), Backus, Chernov, and Zin (2014), Dybvig, Ingersoll, and Ross (1996). Finally, under certain conditions on the pricing kernel process, we show the existence of the {\em long-term short rate}, without stationary and ergodic assumptions on the interest rate process.

{\em Sixth}, we make use of these new long-term measures to several leading asset pricing models. For the first long-term measure, $\pi(m;s)$, we find that the long risk model (Bansal and Yaron (2004)) performs better than the disaster model (Backus, Chernov, and Martin (2011)).  However, with the second long-term measure $z_{\infty}(m)$, the disaster model performs better than the long risk model. Moreover, the internal habit model (Campbell and Cochrane (1991)) is comparable to the disaster model.\footnote{In these models, I only calibrate the standard long risk model and standard disaster model without considering some extensions of these models with complicated statistics components. Therefore, our comparison is not complete to judge these leading asset pricing models. Instead, our comparison exercise justifies to some extent the extensions of these models as in recent literature. See Backus, Chernov, and Zin (2014), Collin-Dufresne, Johnnes, and Lochstoer (2016), Kaltenbrunner and Lochstoer (2010),  Liu (2021), and Pohl, Schmedders, and Wilms (2018).}
Overall, our empirical results are consistent with several recent key observations that the conditional variance of the stochastic discount factor should contain some non-stationary and non-linear factors.


The remainder of the paper is structured as follows. We present an additive version of the long-run law of unexpected shocks in Section \ref{sec:additive}. We also introduce several variations of this long-run law in this section. Section \ref{sec:additive-return} presents applications of the long-run law to asset returns. Section \ref{sec:additive-sdf} shows the applications to the stochastic discount factor and risk-adjusted asset returns.  In Section \ref{sec:multiplicative} we present a multiplicative version of the long-run law and characterize the long-term entropy. 
Section 6 concludes, and technical developments are in Appendix. More technical details are given in the Online Appendix.

\section{A long-run law of unexpected shocks}
\label{sec:additive}

This paper considers a discrete-time economy with an infinite time horizon, $t = 1, 2, \cdots$. The state of nature is represented by $(\Omega,  {\cal F}, ( {\cal F}_t) ,P)$, where $ {\cal F}_t$ denotes the set of all available information up to time $t$, ${\cal F} = ({\cal F}_t)$ is a filtration of sigma-algebras ${\cal F}_t$, and $P$ is a probability measure. $\mathbb{E}_{t}[\cdot]$ denotes the expectation  conditional on information available at time $t$ when no misunderstanding may arise.  

An economic variable $Y$ is represented by a ${\cal F}$-adapted process $(Y_t)$. In this framework, an economic variable can be observable such as an asset price, asset price return, consumption (level) or growth rate, interest rate, inflation, weather and mortality data; and it can be also unobservable such as stochastic discount factor, risk-adjusted asset returns and pricing kernel. If $Y$ is observable, we call each $Y_t$ an observation at time $t$. If $Y$ is not observable, $Y_t$ is the realization of the variable $Y$ under certain model assumptions. For a consistent purpose, we name $Y_t$ the {\em realized value} at time $t$. Throughout this paper, the process $(Y_t)$ satisfies the following assumption.

{\bf Assumption I.} For each $t = 1, \cdots $, $\mathbb{E}_{t-1}[Y_t] < \infty, P-a.s.$

Since $\mathbb{E}_{t-1}[Y_t]$ is the best forecasting of $Y_t$ from the perspective of time $t-1$, the forecast is finite by Assumption I. The difference $Y_t - \mathbb{E}_{t-1}[Y_t]$ between the realized value and its forecasting value is the one-ahead forecasting error, representing the {\em unexpected shock} between time $t-1$ and time $t$. In terminology, we do not distinguish it from {\em shock}  or {\em martingale difference} in this paper.

Define a sequence of random variables,
\be
U_n(Y) = \frac{\sum_{t=1}^{n} (Y_t - \mathbb{E}_{t-1}[Y_t])}{n}, n = 1, 2, \cdots,
\ee
as the arithmetic average of all one-ahead forecasting errors up to time $n$. The main result of this section is an asymptotic property of $U_{n}(Y)$ when $n \rightarrow \infty$, a long-run law of the unexpected shocks. 

To guarantee the existence of the long-run law, the following assumption is imposed. 

{\bf Assumption II.}  There exists a positive number $L$ such that $\mathbb{E}[Var_{t-1}(Y_t)] \le L, t = 1, \cdots$.

	\begin{pr}
	\label{pr:additive}
	Under Assumption I and II for $(P, Y)$, then, for any $\epsilon > 0$, 
	\beq
	U_n(Y) = o( n^{- \frac{1}{2} + \epsilon}), P-a.s.
	\eeq
	In particular,
	\beq
	\lim_{n \rightarrow \infty} \frac{\sum_{t=1}^{n} \left( Y_t - \mathbb{E}_{t-1}[Y_t] \right)}{n} = 0, P- a.s.
		\eeq
	 Moreover, if each $Y_t \in L^2(\Omega,P)$, then
	\beq
	\lim_{n \rightarrow \infty} n^{\frac{1}{2} - \epsilon} U_n(Y) = 0,
	\eeq
	in $L^2(\Omega,P)$.
	\end{pr}
	
   
    Assumption II in Proposition \ref{pr:additive} is not only sufficient but also necessary to guarantee the long-run law in general.
 To demonstrate it, let $Y_n = n \zeta_n$, where $\zeta_n$ are IID, ${\cal N}(0,1)$. Then $\mathbb{E}[Var_{n-1}(Y_n)] = n^2 \rightarrow \infty$. Since $\frac{\zeta_1 + \cdots +  n\zeta_n}{n} \sim N(0, \frac{(n+1)(2n+1)}{6n})$ and $\frac{(n+1)(2n+1)}{6n} \rightarrow \infty$, the sequence of normal random variables $\frac{\zeta_1 + \cdots +  n\zeta_n}{n}$ does not converges to zero almost surely. Actually,  this sequence does not converge to any random variable almost surely since a limit of normal random variables is a normal random variable. Moreover, the central limit theorem implies that the number $\frac{1}{2}$ in Proposition \ref{pr:additive} is the best possible exponent.%

 Assumption II is used to relax the stationary and ergodic assumptions in ergodic theory or numerous technical conditions in the strong law of large numbers. We use two examples to illustrate the long-run law in a nonstationary and nonergodic setting.

\begin{exa}
	Assume the unexpected shock $\epsilon_t$ satisfies
	\beq
	\epsilon_t = \sigma_t z_t; \sigma_t^2 = \omega + \beta \sigma_{t-1}^2 + \alpha \epsilon_{t-1}^2, t\ge 1
	\eeq
	where $z_t$ are IID with $\mathbb{E}[z_t]=0, \mathbb{E}[z_t^2]=1$ and given $\sigma_0 > 0, \epsilon_0 = 0$.Then $Var_{t-1}(\epsilon_t) =\sigma_t^2$. Assumption II holds if and only if $|\beta| < 1$. There are several ways to extend this standard GARCH(1,1) model in a nonergodic setting (See Kristensen (2009) for characterization of GRCH(p,q) model). For example, $\epsilon_t = \sigma_t z_t$ and $z_t \sim N(0, \nu_t^2)$ and $z_t$ is independent from ${\cal F}_{t-1}$. In this case, the conditional variance of the shock is $Var_{t-1}(\epsilon_t) = \sigma_t^2 \nu_t^2$. Assumption II holds as long as the variance of $z_t$ is uniformly bounded from above and $|\beta| < 1$. As another example, let $\epsilon_t = \sigma_t z_t \omega_t$, here $z_t$ are IID with $\mathbb{E}[z_t]=0, \mathbb{E}[z_t^2]=1$, but $\omega_t$ is independent from the sigma algebra generated by $\{ {\cal F}_{t-1}, z_t\}$ and $\mathbb{E}_{t-1}[\omega_t^2] \le L, \forall t$.
In each situation, $(\epsilon_t)$ is nonergoric but Assumption II holds, yielding the long-run law in Proposition \ref{pr:additive} for the shocks.
\end{exa}

\begin{exa}
Consider a Bayesian learning model for $(Y_t)$ with a predictive  distribution $p(Y_t|\theta_t)$ for a stochastic and unknown variable $\theta_t$. Assuming $p(\theta_t|Y_t)$ is known, and ${\cal F}_t$ is generated by $\{Y_1, \dots, Y_t\}$, we obtain the posterior dsitribution $Y_{t+1}|{\cal F}_t = f({\cal F}_{t}) + \epsilon_{t+1}$. In some nonergodic settings with hidden, unknown parameters, the variance of the shock does not necessarily converge to zero but moves inside a finite range (Weitzman (2007), Bakshi and Skoulakis (2010)). In this case, Assumption II holds, and the arithmetic average of shocks converges to zero.
\end{exa}

It should be emphasized that the uniform upper bound condition of the unconditional variance in Assumption II is nothing about the convergence of the  conditional variance of the shocks. Clearly, Assumption II does not imply the convergence of the conditional variance. Moreover, Assumption II could fail even though the conditional variance converges to zero almost surely. For  example,  let  $\epsilon_t = \sigma_t  z_t, $ and $\sigma_t^2 = 3 \sigma_{t-1}^2 z_{t-1}^2$. We assume that $z_t \sim N(0,1)$. In this case,  $Var_{t-1}(\epsilon_t) = \sigma_{t}^2$, and $\mathbb{E}[\sigma_t^2] = 3^t \rightarrow \infty$; however, $Var_{t-1}(\epsilon_t)  \rightarrow 0, P-a,s.$ (Nelson (1990)).

\subsection{Alternative versions}

This subsection presents several alternative version of the long-run law of shocks. 


\begin{pr}
	\label{pr:additive-weighted}
 Let $ 0< a \le 1$, under Assumption I and Assumption II, the following equation holds.
	\be
	\label{eq:fad}
	\lim_{n \rightarrow \infty} \frac{\sum_{t=1}^{n} a^{n-t} (Y_t - \mathbb{E}_{t-1}[Y_t])}{n} = 0, P-a.s.
	\ee
\end{pr}


Proposition \ref{pr:additive-weighted} states that the weighted-average of unexpected shocks converges to zero, when a higher weight is associated with a later (closer) sample data. Nagel and Xu (2021) demonstrate implication of forming belief with higher weight to more recent observation (See Section 3 for the application of the long-run law to form expectation). 


The next one is a long-run law of unexpected shock under higher moments.

\begin{pr}
	\label{pr:additive-higher} Assume that $\mathbb{E}\left[ (Y_t - \mathbb{E}_{t-1}[Y_t])^4 \right] \le L, \forall t$. Then, for any $\epsilon > 0$,
	\be
	U_n(Y) = o(n^{-\frac{1}{4} + \epsilon})
	\ee
	in $L^4(\Omega,P)$.
\end{pr} 


Proposition \ref{pr:additive-weighted} - Proposition \ref{pr:additive-higher} are  useful to distinguish the long-run law with the martingale central limit theorem (MCLT). Under certain conditions of truncated dependable variables' conditional moments, the martingale central limit theorem states that $\sqrt{n} U_{n}(Y)$ converges to a normal distribution in probability (See, for instance, Helland (1982, Theorem 2.5 (a) - (c))). These conditional moments condition, however, are stronger than Assumption II and hard to be verified in most situations. Moreover, Assumption II requires a uniform upper bound of the conditional variance, whereas a lower bound of the conditional variance is also needed in the martingale central limit theorem.\footnote{Specifically, in addition to certain Lindeberg condition for MCLT, the series  $\sum_{t=1}^{\infty} \mathbb{E}[Var_{t-1}(Y_t)]  = \infty$. Clearly, if $0 < l \le \mathbb{E}[Var(Y_t)]  \le L, \forall t$ for two positive numbers $l$ and $L$, this condition for MCLT and Assumption II are satisfied. But the condition for MCLT rules out the case that  $\mathbb{E}_{t-1}[Var(Y_t)] $ decays fast in the long run, and a fast decay rate of the unconditional variances of shocks lead to a better convergence rate of the long-run law.} Therefore, Proposition \ref{pr:additive} holds although the corresponding martingale central limit theorem fails. More importantly, there is no counterpart of Proposition \ref{pr:additive-weighted} -  Proposition \ref{pr:additive-higher} in the martingale central limit theorem literature.

	 		 
%


\begin{pr}
\label{pr:martingale}
 Under Assumption I and II for a process $(Y_t)$,
\begin{enumerate}
	\item 
	if $(Y_t)$ is a submartingale (resp. supermartingale), then $\liminf_{n \rightarrow \infty} \frac{Y_n}{n} \ge 0$ (resp. $\limsup_{n \rightarrow \infty} \frac{Y_n}{n} \le 0$), almost surely;
	\item if $(Y_t)$ is martingale, then, for any positive number $\epsilon$,
	\be
	\label{eq:martingale}
	\frac{Y_n}{n} = o(n^{-\frac{1}{2} + \epsilon}).
	\ee
\end{enumerate}
\end{pr}

 
 
For a general process $(Y_t)$, the (additive) Doob-Meyer decomposition is $Y_t = M_t + A_t$ for a martingale component $(M_t)$ and a predictable component $(A_t)$. Since $(A_t)$ is predictable, the unexpected shock in $(Y_t)$ is derived from the martingale  ({\em permanent}) component. Then, Proposition \ref{pr:martingale} implies (if at least one limit exists)
\be
\label{eq:predictable}
\lim_{n} \frac{Y_n}{n} = \lim_{n} \frac{A_n}{n},
\ee
in which the predictable component $(A_n$) determines the long-term rate of $(Y_t)$. 

%
%

Given a general non-stationary process $(Y_n)$, besides the Doob-Meyer decomposition, there are a number of ways to identify shocks with permanent  martingale component. See, for instance, Beveridge and Nelson (1981) and Hansen (2012). In a Markov environment with state variable $(X_t)$ and under certain conditions, Hansen (2012, Theorem 3.1) shows that
\be
\label{eq:BD}
Y_n = \nu n + M_n + g(X_n) - g(X_0)
\ee
where $(M_t)$ is a martingale permanent component and the second component, $g(X_n) - g(X_0)$, is stationary. It is shown that $M_n$ dominates the fluctuation of $Y_n$ over long time horizons. The number $\nu$ represents the trend of the time series data $Y_n$. By Proposition \ref{pr:martingale}, 
\be
\label{eq:trend}
\lim_{n}\frac{Y_n}{n} = \nu + \lim_{n \rightarrow \infty} \frac{g(X_n)}{n}.
\ee
Equation (\ref{eq:predictable}) and (\ref{eq:trend}) demonstrate the difference between these two  martingale decompositions of a process $(Y_n)$. The stationary component in the second martingale decomposition (\ref{eq:BD}) is not predictable, but there are nice statistical properties (Hansen, Heaton and Li (2008), Hansen (2012)). Moveover, Birkhoff's ergodic theorem derives the existence of  $\lim_{n \rightarrow \infty} \frac{g(X_n)}{n}$, thus the long-term rate of $Y$. In contrast, the martingale component in the Doob-Meyer decomposition enables us to characterize the unexpected shocks, $Y_t - \mathbb{E}_{t-1}[Y_t] = M_t - \mathbb{E}_{t-1}[M_t]$. Hence, by Proposition \ref{pr:martingale}, the long-term rate of $Y$ exists if $\lim_{n \rightarrow \infty} \frac{A_n}{n}$ exists, under a  different set of conditions from Hansen  (2012). Moreover, the linear trend parameter $\nu$ in the long run can be revealed from the predictable component $(A_n)$. 


%

	To proceed, we use a number of conventions to keep the notation consistently in applications. (i)  $R_{t,t+1}$ or $R_{t+1}$ denote any risky asset's growth return over the period  $t$ to $t+1$, and $R_{f,t}$ the risk-free rate of growth return over the same time period. The risky asset can be an equity, equity index or a portfolio. In general, $R_{t, t+n}$ is the grown return over an $n$-period $t$ to $t+n$. (ii) $(M_t)$ denotes a pricing kernel (or state price density) process with $M_0 = 1$, and the financial market might be incomplete. Similarly, $m_t = \frac{M_t}{M_{t-1}}, t \ge 1$ denotes the stochastic discount factor over the period $t-1$ to $t$. (iii) $D(t,s)$ denotes the time-$t$ price of a zero coupon bond maturing at $s$. The continuously compounding yield at time $t$ to time $s$ is defined by $y_{t}^{s} = -\frac{log D(t,s)}{s-t}, s > t$. The continuously compounding short rate at time $t$ is written as $r_{f,t} = -log D(t, t+1)$. (iv) Finally, a continuously compounding return of a risky asset over the period $t$ to $t+1$ is written as $r_{t+1} = log R_{t+1}$.

\section{Implications to asset returns}
\label{sec:additive-return}

This section presents several implications, in the form of ``{\em corollaries}", of the long-run law of unexpected shocks to asset returns in a financial market from an asymptotic perspective. 

\subsection{Long-term expected return  and long-term sample mean}

We start with a reformulation of Proposition \ref{pr:additive} as follows.

\begin{cor}
\label{cor:expectation}
Under Assumption I and II for a return process $(R_t)$ and a probability measure $P$, then
\beq
\lim_{n \rightarrow \infty} \left( \frac{ \sum_{t=1}^{n} R_t}{n} - \frac{\sum_{t=1}^{n} \mathbb{E}_{t-1}[R_t]}{n} \right) = 0, P-a.s.
\eeq
\end{cor}

In spite of its innocuous restatement of Proposition \ref{pr:additive}, Corollary \ref{cor:expectation} has interesting implication for asset pricing. In the above expression, the firm term $\frac{R_1 + \cdots + R_n}{n}$ is the average of the realized sample data which is available for a long-lived agent, so it is termed as a sample mean. Its limit (if exists) is a {\em long-term sample mean}.
On the other hand, the second term $\frac{\sum_{t=1}^{n} \mathbb{E}_{t-1}[R_t]}{n}$ depends on the probability measure (belief) $P$ and the distribution (model) of asset return. To be different, we name it the {\em long-term expected return} under belief $P$ and assumption on the asset return. Corollary \ref{cor:expectation} states that the long-term expected return under any belief and model assumption equals the long-term sample mean. 

\begin{cor}
\label{cor:premium}
Under Assumption I and II for a return process $(R_t)$ and a probability measure $P$, then (if at least one limit exists)
\beq
\lim_{n \rightarrow \infty} \frac{ \sum_{t=1}^{n} ( \mathbb{E}_{t-1}[R_{t}] - R_{f, t-1})}{n} = \lim_{n \rightarrow \infty} \frac{ \sum_{t=1}^{n} (R_{t} - R_{f, t-1})}{n}, P- a.s.
\eeq
\end{cor}

Corollary \ref{cor:premium} states that the long-term expected excess return equals the {\em long-term sample excess return},  $\lim_{n \rightarrow \infty} \frac{ \sum_{t=1}^{n} (R_{t} - R_{f, t-1})}{n}$, regardless of the model assumption about $\mathbb{E}_{t-1}[R_t]$.  From an empirical perspective, the sample (arithmetic) average process displays a better stable shape than an asset return process $(R_t)$. For instance, the standard deviation of the sample (arithmetic) average of excess returns is 0.32\% for daily return (from 1962 to 2020), and 1.11\% for monthly return (from 1926 to 2020), respectively, yielding the existence of a long-term sample excess return (see the details in the Online Appendix). 

%

A long-lived agent is able to compute the long-term sample excess return; then, she can use Corollary \ref{cor:premium} to see whether a model is meaningful concerning on the expected return. To estimate a model-free long-term expected return, we follow Martin (2017) to use a model-free lower bound of the expected return with available derivative (S \&P 500 index options) data. Specifically, under Martin (2017)'s negative correlation condition (NCC), and let $Q$ be a risk-neutral probability measure, Martin (2017) shows that $\mathbb{E}_{t-1}[R_t]$ is bounded from below by $R_{f, t-1} + \frac{1}{R_{f,t-1}} Var_{t-1}^{Q}(R_{t})$.\footnote{Here, I employ the model-free expression of $Var_{t}^{Q}(R_{t+1})$ in terms of index call option for the market index $R$ as follows,
\beq
\frac{1}{R_{f,t}} Var_{t}^{Q}(R_{t+1}) = \frac{2}{S_{t}^2} \left\{ \int_{0}^{F_{t, t+1}} Put_{t, t+1}(K)dK + \int_{F_{t, t+1}}^{\infty} Call_{t, t+1}(K) dK\right\}
\eeq
where $S_t$ denotes the index price at $t$, $F_{t, t+1}$ is the index's future value at time $t$ with maturity $t+1$, and Call or Put represent the index call or put option. We follow the same method in Martin (2017) to compute the integrals on the right hand using market available index options.}
Hence, under Assumption I and II for $P$ and $(R_t)$,  but without model assumption about the asset return though, NCC and Corollary \ref{cor:premium} imply
\be
\label{eq:NCC}
\lim_{n \rightarrow \infty} \frac{ \sum_{n=1}^{n} \frac{1}{R_{f,t-1}} Var_{t-1}^{Q}(R_{t})}{n} \le  \lim_{n \rightarrow \infty} \frac{ \sum_{t=1}^{n} (R_{t} - R_{f, t-1})}{n}.
\ee


While Equation (\ref{eq:NCC}) can be verified by a long-lived agent, a short-lived agent is only able to approximate it by a large sample of available data. Still, Equation (\ref{eq:NCC}) is useful with available sample data. As an illustration, Figure \ref{fig:ncc} (the top panel) displays both sides of Equation (\ref{eq:NCC}) approximately by using market available data (index and index options). We use the daily-frequent data over the period 1996 to 2020. Overall, both sides of Equation (\ref{eq:NCC}) are very close to each other, whereas the sample average of excess returns is below the sample average of risk-neutral variances in certain periods after 2001. It suggests that the NCC is barely reasonable asymptotically but not necessarily true in the long run. Bakshi, et.al. (2020) provide some examples of an economy in which the NCC is not supported. 

%
%
%

\subsection{Formation expectation}
So far, we do not discuss the role of the probability measure $P$ in the long-run law. A decision-maker can form (subjective) probabilistic expectations by using historical data (empirical probability)  or survey respondent probability (Manski (2001)). Understanding belief and subjective expectation formation from data have been attracted lots of interest recently in asset pricing. In this subsection, we study the difference between the subjective and objective expectation in a long run. 

Let $\tilde{P}$ and $\tilde{E}$ represent the subjective probability and the corresponding expectation. By contrast, the objective probability is denoted by $P$. 
The next result builds a link between the subjective expectation and objective expectation in the long run as follows. 

\begin{cor}
\label{cor:formation}
Under Assumption I and II for $(P, R_t)$ and $(\tilde{P}, R_t)$, then 
\beq
\lim_{n \rightarrow \infty} \frac{ \sum_{t=1}^{n} ( \mathbb{E}_{t-1}[R_t] - \tilde{E}_{t-1}[R_t]) }{n} = 0, a.s.
\eeq
\end{cor}

This corollary follows directly from Proposition \ref{pr:additive} by comparing with the realized return $R_t$ on each term. It states that the long-run expectation difference between any subjective expectation and objective expectation is zero, as long as these formation expectations do not move too significantly (Assumption II holds for both $P$ and $\tilde{P}$). Malmendier and Nagel (2011, 2016) demonstrate the difference in inflation between the subjective and objective perspectives. Empirically, Nagel and Xu (2021) also demonstrate the difference between subjective asset return and objective asset return.\footnote{Here, we consider the difference between expectations. As will be shown in Section 4, there is an extra {\em convexity} term if we compare the log-expectation between the subjective and objective probability, $log \mathbb{E}_{t-1}[R_t] - log \tilde{E}_{t-1}[R_t]$. See also Nagel and Xu (2021), equation (36) - (37).} However, the difference between the long-term subjective expectation and the long-term objective expectation of an economic variable should be merely small and disappears in a long run. Furthermore, the long-term expected excess return is independent of the subjective probability (belief).

\subsection{Risk-neutral expectation}

Since the long run law of unexpected shocks in Proposition \ref{pr:additive} holds for any probability measure, it is natural to consider the risk-neutral probability measure in Proposition \ref{pr:additive}, assuming the existence of a risk-neutral probability measure $Q$ in a financial market.

\begin{cor}
\label{cor:risk-neutral}
Let $Q$ be a risk-neutral probability measure, and for one risk asset with return process $(R_t)$, there exists one positive number $L$ such that
\be
\label{eq:rn}
Var^{Q}(R_{t+1}) - Var^{Q}(R_{f,t}) \le L, \forall t = 1, 2, \cdots,
\ee
then the long-term sample excess return of this asset is {\em zero}.
\end{cor}

\begin{figure}
\begin{center}
\caption{Arithmetic average of risk-neutral variance (derivative) and risk premium}
\label{fig:ncc}
\rule{0pt}{3pt}
\parbox{7in}
\medskip
\parbox{6.5in}
{\small The top pannel displays the daily time-series risk premium (in annual) of S \P 500 index over the risk-free rate of return from Jan 4,1996 to Dec 31, 2020, and the time-series risk-neutral variance, 
$\frac{\sum_{n=1}^{n} \frac{1}{R_{f, t-1}} Var^{Q}_{t-1}(R_t) }{n}$ in the same time period. Under Negative Correlation condition (between asset return and its risk-adjusted asset return), Martin (2017) shows that the expected risk premium is bounded from below by $\frac{1}{R_{f, t-1}} Var^{Q}_{t-1}(R_{t})$.  
The bottom panel displays the time series of VIX (in percent)  in the same time period. The VIX is divided by 10 to have a better comparison with the risk-neutral variance on level. The correlation between the risk-neutral variance and VIX is 0.84, so the risk-neutral variance is also a reasonable measure of the financial market turmoil.
}
\bigskip
\includegraphics[width=6in, height=2.5in]{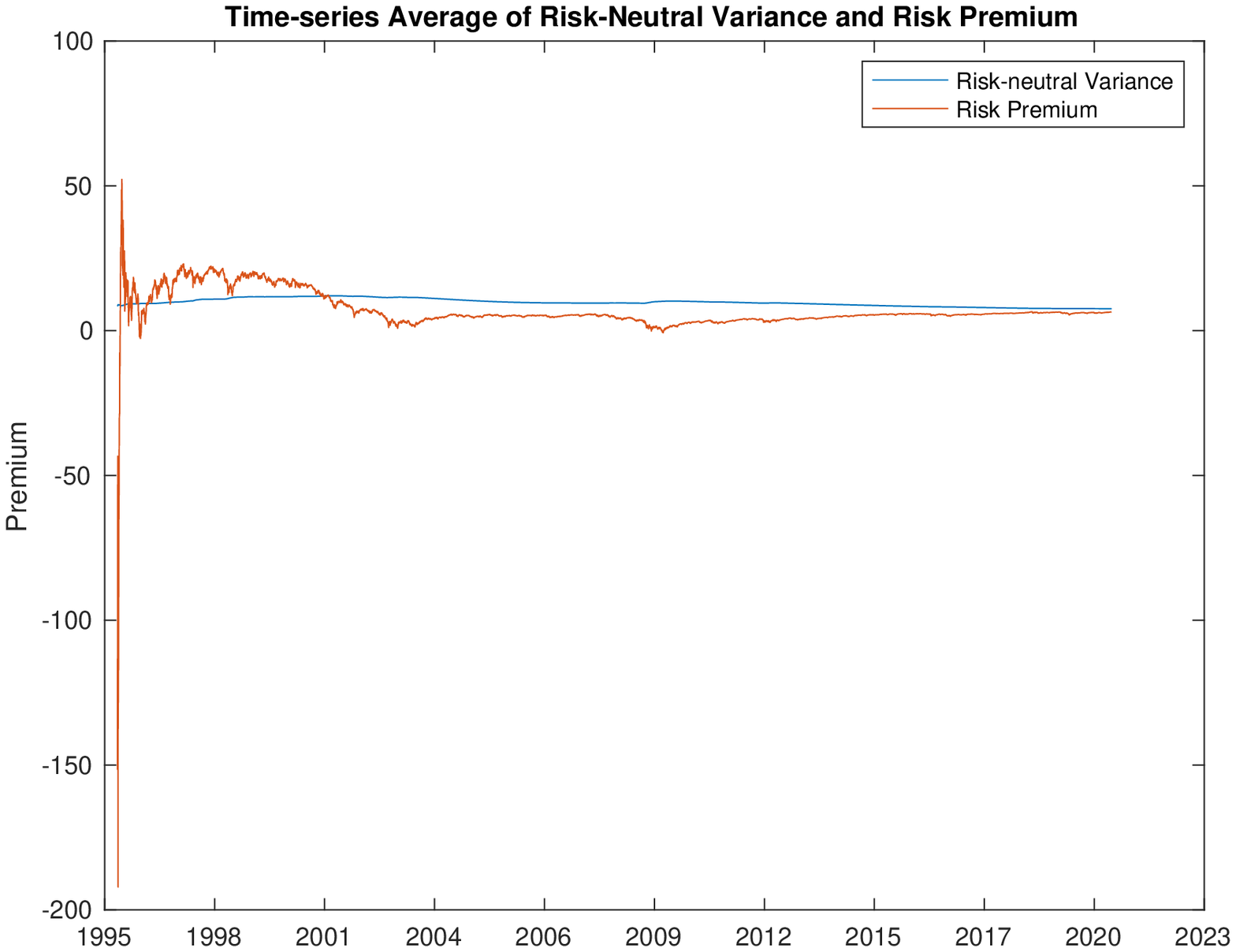}
\includegraphics[width=6in, height=2.5in]{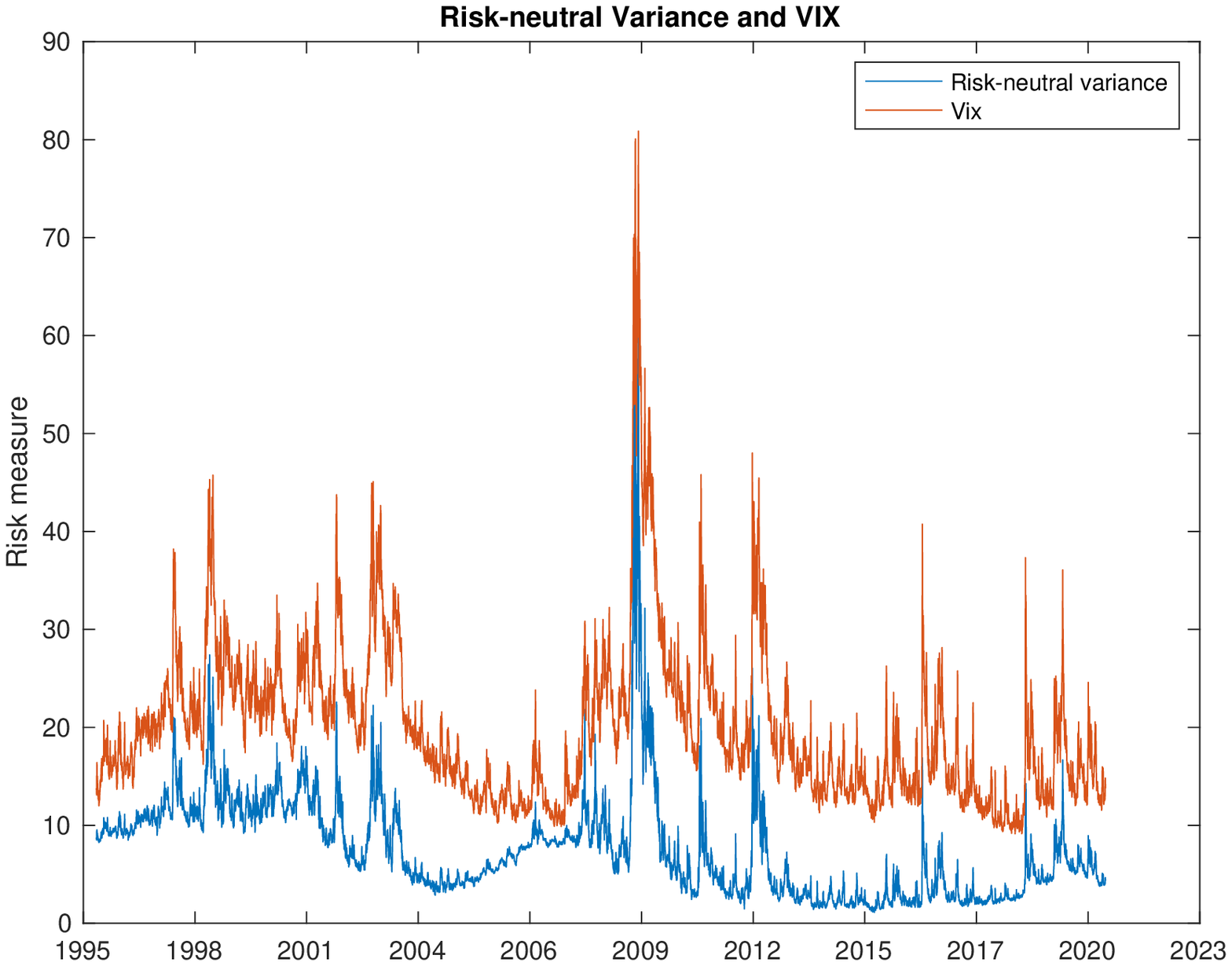}
\end{center}
\end{figure}

Under the risk-neutral probability measure $Q$, Assumption I is evident as $\mathbb{E}^{Q}_{t}[R_{t+1}] = R_{f,t} < \infty$. Since Assumption II folds for $(Q, (R_t))$ by Equation (\ref{eq:rn}),  Corollary \ref{cor:risk-neutral} follows directly from Proposition \ref{pr:additive}  and Corollary \ref{cor:premium} (for the risk-neutral measure). However, Corollary \ref{cor:risk-neutral} seems counterintuitive for the following reason.
Let us consider the market (index) return $R$ as an example. On the one hand, the bottom panel of Figure \ref{fig:ncc} plots the time series of $Var_{t}^{Q}(R_{t+1})$ between 1996 to 2020. In average, the level of risk-neutral variance is about 1.55 percent, and takes only significant value at certain time period. 
Therefore, it is reasonable to argue that Equation (\ref{eq:rn}) holds for the market return. On the other hand, it is also empirically solid that the long-term sample excess return of the market index is positive (positive equity premium). Granted, Corollary \ref{cor:risk-neutral} implies that there exists {\em free lunch} in the market since there is {\em no} risk-neutral probability measure!

\begin{exa}
\label{exa:risk-neutral} Consider a financial market with one risky asset (index) and its return process under a probability measure $P$ satisfies $R_t = R_{t-1} + \epsilon_t + \alpha_t$, where $(\epsilon_t)$ is a Rademacher sequence of independent random variables $\epsilon_i$ with 
\beq
P(\epsilon_t = +1) = P(\epsilon_t = -1) = \frac{1}{2},
\eeq
 and $\{ \alpha_t\}$ is a deterministic sequence of real numbers in $(0, 1)$. Assuming  $R_0 = 0$ and the rate of risk-free interest is always zero. It is straightforward to see that $Q$ is an unique martingale measure if and only if \beq
 Q(\epsilon_t  =+1) = \frac{1- \alpha_t}{2}, Q(\epsilon_t = -1) = \frac{1+ \alpha_t}{2}.
 \eeq
 Therefore, $Var^{Q}(\epsilon_t) = 1-\alpha_t^2$ and $Var^{Q}(R_t) = \sum_{j=1}^{t}(1-\alpha_j^2)$. 
\end{exa}

In this example of Schachermayer (1994), Assumption II holds for $(Q, (R_t))$ if and only if the series $\sum_{n=1}^{\infty} (1-\alpha_n^2)$ is finite. By Kakutani's theorem (see, Willams (1991, 12.7))\footnote{This theorem of Kakutani is used in Martin (2012)'s long-dated asset valuation  theory. See Section 4.3 below.}, this infinite series is finite if and only if $Q$ is equivalent to $P$. Put it differently, if the series $\sum_{n=1}^{\infty} (1- \alpha_n^2) = +\infty$, then $Q$ and $P$ are mutually singular; and therefore, there is no equivalent martingale measure in this financial market. As a consequence, there is a free lunch with bounded risk. Hence, for this example, Assumption II holds if and only if there is no arbitrage opportunity.

 Corollary \ref{cor:risk-neutral} is useful to explain the  pervasive financial market meltdown phenomenon in the equity market from a long-lived agent's perspective. We again consider the equity market index return and build the following discussion on Martin (2017). The bottom panel of Figure \ref{fig:ncc} displays a substantial similarity between the risk-neutral variance of the market index and VIX in the market. For instance, the correlation between the risk-neutral variance and VIX is above 0.83. Therefore, a higher value of VIX is associated with the risk neutral variance. The risk-neutral variance is, in essence, the SVIX index introduced in Martin (2017), and Martin (2017) demonstrates the SVIX is also a good measure of equity market turmoil and financial crisis in general. For instance, the VIX takes spike value 80-85 during 2008 financial crisis and 2020 Covid-19 period. That is to say, an extremely large value of the risk-neutral variance of the market (index) return is fairly consistent with the financial market crisis.
 
 \begin{df}
 	There is a {\em financial market crisis} in the period $t-1$ to $t$ if $Var_{t-1}^{Q}[R_t]$ is sufficiently large, for the market return process $(R_t)$.
 \end{df}

 \begin{cor}
 	\label{cor:crisis}
Assuming the equity market is no-arbitrage and the positive long-term sample excess return of the (equity) market index, then for any positive number $L$ there exists some future time $t$ such that $\mathbb{E}^{Q}\left[Var_{t-1}^{Q}(R_{t})\right] > L$, thus, there must have financial  market crisis persistently.
\end{cor}

According to Corollary \ref{cor:crisis}, any long-lived agent must see either {\em arbitrage opportunities} in the equity market or significant {\em equity market turmoil} persistently. 


\subsection{An application to survey expectation}

This subsection presents an application to the survey expectation that whether this survey return reflects a risk-neutral expectation or a pessimistic expectation return. Adam, Matveev, and Nagel (2021) demonstrate that both hypotheses are wrong empirically and robustly.  As an application of the long-run law, we provide an alternative theoretical argument for why these hypotheses are invalid since the long-term sample excess return of the market (or any risky asset) is positive.

For any agent $i$ with a subjective probability measure $P^i$, $Q^i$ represents her martingale measure. Here, we only use the martingale measure, not a stronger risk-neutral measure condition, and do not need equivalence between each subjective probability measure. Actually, these subjective probability measure can be mutually singular.\footnote{Notice that the long-term sample excess return is defined by the same {\em observable} sequence $\frac{(R_1 - R_{f, 0}) + \cdots + (R_n - R_{f, n-1})}{n}$ for all agents. If $P^i$ are equivalent, then a positive long-run excess return sample mean for one agent implies the same property for all agents. In its general expression, we highlight particular agent $i$ as stated in Corollary \ref{cor:hypothesis}. }  $\tilde{E}^i[\cdot]$ denotes the subjective expectation under $P^i$.

Following Adam, Matveev, and Nagel (2021), write the following risk-neutral hypothesis and pessimistic hypothesis, respectively\footnote{See Adam, Matveev, and Nagel (2021), equation (3) and equation (10). Notice that a pessimistic expectation follows from a negative correlation between the agent's marginal-utility and the return.},
\be
\text{ (Risk-Neutral) \ } \tilde{E}^{i}_{t}[R_{t+1}] = \mathbb{E}_{t}^{Q^i}[R_{t+1}] + \epsilon^{i}_t,
\ee
and
\be
\text{(Pessimistic) \ } \tilde{E}^{i}_{t}[R_{t+1}] < \mathbb{E}_{t}^{Q^i}[R_{t+1}] + \epsilon^{i}_t
\ee
where the measurement error $\epsilon^{i}_t$  captures the fact that the agent empirically measure expectations with noise. Assume each noise $\epsilon^{i}_t \in {\cal F}_t$ and $\tilde{E}^{i}_{t-1}[\epsilon^{i}_t] = 0$, and the variance of $\epsilon_{t}$ are uniformly bounded from above for all $t$. 

\begin{cor}
\label{cor:hypothesis} Under Assumption I and II for $(P^i, R_t)$, and the long-term excess return for $(R_t)$ is positive for agent $i$, then Risk-Neutral and Pessimistic Hypothesis fail for agent $i$.
\end{cor}

\section{Implication to stochastic discount factor (SDF)}
\label{sec:additive-sdf}

This section discusses the application of the long-run law to the pricing kernel process and the risk-adjusted asset return processes. The financial market is non-arbitrage. There exists a pricing kernel or state-price density process $(M_t)$ with a given probability measure $P$.

\subsection{Long-term SDF}


 We start with the pricing kernel process. Since $\mathbb{E}_{t}[m_{t+1}] = D(t, t+1)$, the next result follows from Proposition \ref{pr:additive}.

\begin{cor}
\label{cor:pricing-kernel-additive}
If there exists a positive number $L$ such that $Var(m_{t}) \le Var(D(t-1,t)) + L$, then
\beq
\lim_{n \rightarrow \infty} \left( \frac{m_1 + \cdots + m_n}{n}  - \frac{D(0,1) + \cdots + D(n-1, n)}{n}\right) = 0,  P-a.s.
\eeq
Moreover, the converges holds in $L^2$.
\end{cor}

Here, Assumption II for the stochastic discount factor follows from a uniform bound condition of the unconditional variance of the stochastic discount factor $m_t$. Therefore, Bekaert and Liu's (2004)  bounds of $Var(m_t)$ can be used to verify Assumption II using the first and second moment of basic asset payoffs. Moreover, it is equivalent to the upper bound on the Sharpe ratio of the portfolio (Hansen and Jagannathan (1991)). In this respect, Assumption II is reasonable from the no-arbitrage perspective since it relates to the good-deal bounds in Cochrane and Saa-Requejo (2000). 

 Under a stationary and ergodic assumption on the short rate process $\{ r_{f, t}\}$, it is straightforward to derive the existence of $\overline{d} =  \lim_{n \rightarrow \infty} \frac{D(0,1) + \cdots + D(n-1,n)}{n}$.\footnote{It is possible to construct non-ergodic term structure model. For example in Ingersoll, Skelton, and Weil (1978), $r_{f,t} = r_0 + \delta N_t$ where $N_t$ is a Poisson process with intensity $\lambda$ and jump size 1. A sequence $(a_n)$ might diverge but its arithmetic average (Cesaro sum) converges. For example, $a_n = 1$, for even $n$ and $a_n = 0$ for odd $n$. Then $\lim_{n \rightarrow \infty} \frac{a_1 + \cdots + a_n}{n} = \frac{1}{2}$.} Corollary \ref{cor:pricing-kernel-additive} demonstrates that the avergae of SDFs ( long-term SDF), $\frac{m_1 + \cdots + m_n}{n}$, is sufficiently close to the sample average of the short-term bond prices, $\frac{D(0,1) + \cdots + D(n-1, n)}{n}$, regardless of the ergodic assumption of the short rate process or not. 

\subsection{Volatility of SDF}

In this subsection, we derive a long-run property of higher moments of pricing kernels, building on the nonparametric bound literature of the stochastic discount factor (Hansen and Jagannathan (1991), Snow (1991), Bansal and Lehmann (1997), Alvarez and Jermann (2005), and Liu (2021)). Building on this property, we next introduce a long-term measure of the stochastic discount factor to compare several leading asset pricing models and discuss its applications.

%

\begin{cor}
\label{cor:liu}
Let  $s$ be a real number, $s \ne 0, s \ne 1$,  assume Assumption I and Assumption II hold for $(P, (m_t^{s}))$. Then, for either $s > 1$ or $s < 0$,
\be
\liminf_{n \rightarrow \infty} \frac{ \sum_{t=1}^{n} m_{t}^{s}}{n} \ge \limsup_{n \rightarrow \infty} \frac{ \sum_{t=1}^{n} \mathbb{E}_{t-1}\left[ (R_{t})^{-\frac{s}{1-s} }\right]^{1-s} }{n}
\ee
where $R = (R_t)$ runs through all asset return processes. If $0 < s < 1$, then
\be
\limsup_{n \rightarrow \infty} \frac{\sum_{t=1}^{n} m_{t}^{s}}{n} \le \liminf_{n \rightarrow \infty} \frac{ \sum_{t=1}^{n} \mathbb{E}_{t-1}\left[ (R_{t})^{-\frac{s}{1-s} }\right]^{1-s} }{n}.
\ee
\end{cor}

Corollary \ref{cor:liu} is closely related to Snow (1991) for $s > 1$ and Liu (2021) for $s < 1$. 
It states that the  time series of higher moments of stochastic discount factors are bounded (from above or below) by the arithmetic average of the higher-order moments of asset returns.

By Corollary \ref{cor:liu}, it is temping to introduce a measure (assuming the limit exists), $
\lim_{n \rightarrow \infty} \frac{\sum_{t=1}^{n} m_{t}^{s}}{n}$, 
to diagnose asset pricing models. To make good use of this measure, we need to estimate both the stochastic discount factors and higher-order moments of asset return. For the first one, a specification of a pricing kernel is often derived from representative agent's preference and macro-economic data like consumption growth or market return data. However, to calculate the higher-moments of asset returns is challenge since it depends on distribution assumptions on asset returns. 

To avoiding assumption about asset return distribution, we derive a duality result of Corollary \ref{cor:liu} below, in which the high-order moments of stochastic discount factors and only the {\em sample} data of asset returns are required.


\begin{cor}
\label{cor:liu-dual} Given a real number $s \ne 0, s \ne 1$ and Assumption I and Assumption II hold for $(P, (R_{t}^{\frac{s}{s-1}}))$, then,
\beq
\lim_{n \rightarrow \infty} \frac{ \sum_{t=1}^{n} \mathbb{E}_{t-1}[ m_{t}^s ]^{ \frac{1}{1-s}} }{n} \le \liminf_{n \rightarrow \infty} \frac{ \sum_{t=1}^{n} (R_{t})^{\frac{s}{s-1}}}{n}, \text{ if } s >0,
\eeq
and
\beq
\lim_{n \rightarrow \infty} \frac{\sum_{t=1}^{n} \mathbb{E}_{t-1}[ m_{t}^s ]^{ \frac{1}{1-s}} }{n} \ge \limsup_{n \rightarrow \infty} \frac{\sum_{t=1}^{n} (R_{t})^{\frac{s}{s-1}}}{n},  \text{ if } s < 0.
\eeq
\end{cor}

The crucial point in Corollary \ref{cor:liu-dual} is that there is no model assumptions on the asset returns. Since a long-lived agent knows the realized asset returns from the sample data, she can use Corollary \ref{cor:liu-dual} to check whether an asset pricing model provides appropriate SDF to fit the data. Define
 \be
 \pi(m;s) =\lim_{n \rightarrow \infty} \frac{ \sum_{t=1}^{n} \mathbb{E}_{t-1}[ m_{t}^s ]^{ \frac{1}{1-s}} }{n},
 \ee
 for each pricing kernel process $(M_t)$ and any real number $s \ne 0, s \ne 1$, and name it a {\em long-term higher moments of SDF}.

To illustrate, we make use of two leading asset pricing models in Corollary \ref{cor:liu-dual} and check which model has a better fit with the sample data of historical asset returns. The first one is a disaster model (Barro (2006),  Backus, Chernov, and Martin (2011)). The second one is a long run risk model (Bansal and Yaron (2004)). 

\begin{exa}
The stochastic pricing factor in a disaster model is given by
\beq
m_{t+1} = \beta g_{t+1}^{-\gamma},  log(g_{t+1}) = \epsilon_{t+1} + \eta_{t+1}
\eeq
where $g_{t+1} $ is the consumption growth rate, $\epsilon_{t+1} \sim {\cal N}(\mu,\sigma^2)$, and $\eta_{t+1}| (J = j) \sim {\cal N}(j \theta, j \nu^2)$ and $J$ is a Poisson random variable with the jump intensity parameter $\omega$. 
$\epsilon_{t+1}$ and $\eta_{t+1}$ are independent. 
\end{exa}

 In this disaster model,  for any $s \ne 0$ (see Liu (2021), equation (C5)),
\be
\mathbb{E}_{t}\left[m_{t+1}^s\right] = exp\left\{ s (log \beta - \gamma \mu)  + \frac{1}{2} \gamma^2 \sigma^2 s^2 + \omega[ e^{-\gamma s \theta + (\gamma s \nu)^2/2} - 1 ] \right\}.
\ee
Since $\mathbb{E}_{t}\left[m_{t+1}^s\right]$ is a constant across the time, the left side in Corollary \ref{cor:liu-dual} is calculated from the last equation easily.

\begin{exa}
The stochastic factor in a long run risk model is (in Bansal and Yaron (2004), equation (A2) and equation (3) therein)
\be
m_{t+1}= \delta^{\theta} exp\left\{ ( \theta-\frac{\theta}{\psi} - 1) g_{t+1} + (\theta -1)(\kappa_0 + \kappa_1 z_{t+1} - z_{t}) \right\}
\ee
where $\theta = \frac{1 - \gamma}{1 - \frac{1}{\psi}}, z_t = A_0 + A_1 x_t + A_2 \sigma_{t}^2$ and the state variable $(x_t, g_{t}, \sigma_{t})$ sasisfies
\beq
g_{t+1} = \mu + x_t + \sigma_{t} \eta_{t+1}, x_{t+1} = \rho x_t + \phi_{e} \sigma_{t} e_{t+1}, \sigma_{t+1}^2 = \sigma^2 + \nu_1 (\sigma_{t}^2 - \sigma^2) + \sigma_{w} w_{t+1},
\eeq
and IID $w_{t+1}, e_{t+1},\eta_{t+1} \sim {\cal N} (0,1).$ $\kappa_0, \kappa_1, A_0, A_1$ and $A_2$ are calculated explicitly by model parameters.
\end{exa}

 In this long run risk model, following Bansal and Yaron (2004),
\beq
\mathbb{E}_{t}\left[m_{t+1}^s\right] = exp\{\phi(s) + \alpha(s) \sigma_{t}^2 + \beta(s) x_{t}\}
\eeq
for three deterministic functions $\phi(s), \alpha(s)$ and $\beta(s)$. Since $(\sigma_{t}^2, x_t)$ is stationary and ergodic, $\mathbb{E}_{t}\left[m_{t+1}^s\right]$ is also a stationary and ergodic process. Therefore, it is straightforward to obtain
\be
\pi(m;s) = exp\left\{\frac{\phi(s)}{1-s} + \frac{\alpha(s)}{1-s} \sigma^2  + \frac{\alpha(s)^2}{2 (1-s)^2}  \frac{\sigma_{w}^2}{1-\nu_1^2} + 
\frac{\beta(s)^2}{2(1-s)^2}  \frac{\phi_{e}^2 \sigma^2}{1-\rho^2} \right\}.
\ee

Figure \ref{fig:moments} displays the long-term higher moments of a stochastic discount factor in the disaster and long run risk models. For the disaster model, we use the values of parameters $\{\beta, \gamma, \mu, \sigma^2, \omega, \theta, \nu^2\}$ calibrated in (Liu (2021, Table C.1)) and the Baseline parameters $\omega_B, \theta_B$. In the long run risk model, we use the value of parameters in Bansal and Yaron (2004), Bansal, Kiku, and Yaron (2012). According to Corollary \ref{cor:liu-dual}, for a {\em negative} value of $s$, the new diagnostic tool, $\pi(m;s)$, is a upper bound of long-term higher moments of asset returns. Then, a larger value of the long-term higher moments of $m_{t+1}$ leads to a better asset pricing model. As shown, the long run risk model performs better than the disaster model in using Corollary \ref{cor:liu-dual}. On the other hand, for a {\em positive} value of $s$, since $\pi(m;s)$ becomes a lower bound of long-term higher moments of asset returns, the smaller the value of $\pi(m;s)$ for positive $s$, the better the asset pricing model. In this respect, the long run risk model is also better than the disaster pricing model.

Since $\pi(m;s)$ offers a long-term measure to diagnostic asset pricing models by Corollary \ref{cor:liu-dual} and there is a persistent component in the long run risk model, it is reasonable to expect that the long run risk model is better than the disaster model from a long run perspective. In a disaster model, the higher moments of stochastic discount factors are a constant. There is no conditional variance of the stochastic discount factor. By contrast, there is a persistent, predictable component $x_t$ in the consumption process in the long run risk model. Even though $\mathbb{E}_{t}[m_{t+1}^s]$ is still stationary and ergodic, its conditional variance is determined by the randomness of the process $x_t$ and stochastic variance process $\sigma_{t}^2$. 
Consistent with Backus, Chermov and Zin (2014), and Liu (2021), Figure \ref{fig:moments} suggests the importance of the conditional variance of the stochastic discount factor or higher moments in building asset pricing models.

\subsection{Long-run law of risk-adjusted asset return}

In this subsection, we study the long-run law of the risk-adjusted asset return process. 

Notice that $\mathbb{E}_{t-1}[m_t R_t] = 1$, the next result follows from Proposition \ref{pr:additive} clearly.

\begin{cor}
\label{cor:risk-adjusted}
Assume that $Var(m_t R_t) \le L, \forall t$ for a positive number $L$, then
\be
\frac{m_1 R_1 + \cdots +m_n R_n}{n} = 1+ o(n^{-\frac{1}{2} + \epsilon}), 
\ee
 for any positive number $\epsilon$. 
\end{cor}

Assuming the negative correlation between the pricing kernel with the asset return, the uniform upper bound of the unconditional variance of $m_t R_t$ follows from the uniform upper bound of the variance of the pricing kernel, and the second moment of asset return. If so, the Casero sum of the sequence $m_n R_n$ converges to 1, almost surely. As a consequence, if $m_nR_n$ converges, its limit must be one as well.

 Martin (2012) studies the properties of the positive martingale $(m_1R_1 \cdots m_nR_n)$ when $n$ goes to infinity (long-dated asset pricing). Specifically, assuming  independent risk-adjusted asset return $m_n R_n$, then $m_n R_n$ converges if and only if
$\sum_{n=1}^{\infty} Var(\sqrt{m_n R_n}) < \infty$, and if so, $m_n R_n \rightarrow 1$. On the other hand, if the series $\sum_{n=1}^{\infty} Var(\sqrt{m_n R_n}) = \infty$, it is shown that $\lim_{n \rightarrow \infty} (m_1R_1 \cdots m_n R_n) = 0$.\footnote{By a non-generic case in Martin (2012) it  means that $m_nR_n \rightarrow 1, a.s$.  If there exists a positive number $\delta$ such that the series $\prod \mathbb{E}_{t-1}[\sqrt{m_t R_t}] \ge \delta, a.s.$, then it is shown that $m_nR_n \rightarrow 1, a.s.$. On the other hand, if the product series $\prod \mathbb{E}_{t-1}[\sqrt{m_t R_t}] $ diverges almost everywhere (generic), it can be shown that $m_1R_1 \cdots m_n R_n \rightarrow 0, a.s.$}
Clearly, a uniform bound of the variance of the risk-adjusted asset return (Assumption II) is weak compared with a convergent series of the variances in Martin (2012). Moreover, even though $m_n R_n$ diverges in general, its Cesaro sum converges to 1 with a converge rate $\frac{1}{2} - \epsilon$. 

In contrast to Martin's {\em probabilistic} approach, we next present an {\em analytical} approach to show that $m_nR_n$ diverges in a generic sense, and $m_nR_n$ converges to one in certain special cases. 


\begin{cor}
	
\label{cor:martin-analysis}
Assume that $Var(m_t R_t) \le L, \forall t$ for a positive number $L$.
\begin{enumerate}
	\item If $m_nR_n$ {\em slowly decreases} in the sense that $\liminf (m_kR_k - m_n R_n) \ge 0, a.s.$ when $\frac{k}{n} \rightarrow 1, k > n \rightarrow \infty$, then $\lim_{n \rightarrow \infty} m_n R_n = 1, a.s.$ (non-generic case)
	\item If $\liminf( m_kR_k - m_nR_n) < 0, a.s., $ for certain sequence $\{n, k \}$ such that $\frac{k}{n} \rightarrow 1, k > n \rightarrow \infty$, then $m_n R_n$ diverges, a.s. (generic case)
\end{enumerate}
\end{cor}

By Corollary \ref{cor:risk-adjusted}, the infinite series $\sum a_n$ is Cesaro summable\footnote{A series $\sum_{n=1}^{\infty}x_n$ of real numbers $x_n$ is classical summable if the partial sum $s_n = x_1 + \cdots + x_n$ has a finite limit wnen $n$ goes to infinity. It is Cesaro summable if the sequence $\frac{s_1 + \cdots s_n}{n}$ has a finite limit when $n$ goes to infinity.}, where $a_n = m_n R_n - m_{n-1}R_{n-1}$. In analysis, to show one Cesaro summable series $\sum a_n$ is summable under certain conditions is the classical Tauberian theory (Korevaar, 2004). For instance, in Corollary \ref{cor:martin-analysis}, if the risk-adjusted asset short-term return $m_n R_n$ in the time period $[n , n+1]$ slowly decreases, then $m_nR_n$ converges to its Casero limit, 1. 

Nevertheless, due to the high degree of uncertainty of the risk-adjusted return, the risk-adjusted short-term return $m_nR_n$ does not {\em slowly decrease} in general. To illustrate, the condition that the risk-adjusted return in the period  $[k-1,k]$ is strictly smaller than the risk-adjusted return in the period $[n-1,n]$ denotes that there is a {\em reversal} of the risk-adjusted return from the period $[k-1,k]$ to the period $[n-1,n]$. For a long-lived asset, there should be infinitely many reversals of the risk-adjusted return; otherwise, the risk-adjusted return would increase eventually, a contradiction to the risky nature of the financial asset return. Therefore, the sequence of the risk-adjusted asset returns $m_nR_n$ should diverge in the generic case.

\section{A multiplicative theory and the long-term entropy}
\label{sec:multiplicative}

This section develops a theory of the multiplicative version of the long-run property of economic variables and presents its implications to asset returns and stochastic discount factor. Given a non-negative process $Y_t$ with $\mathbb{E}_{t-1}[Y_t] >0, a.s.$, the multiplicative version of $U_n(Y)$ is
\be
V_n(Y) = \left( \prod_{t=1}^{n} \frac{Y_t}{\mathbb{E}_{t-1}[Y_t]} \right)^{\frac{1}{n}}, n = 1, 2, \cdots.
\ee
For any ${\cal F}$-adapted process $(x_{t})$,  the conditional entropy
\be
J_t(x_{t+1}) = log \mathbb{E}_{t}[x_{t+1}] - \mathbb{E}_{t}[log(x_{t+1})] \ge 0.
\ee
This conditional entropy measures the risk of the ${\cal F}_{t+1}$-variable $x_{t+1}$. The entropy process $(J_{t}(x_{t+1}))$ measures the dynamic risk of the process $(x_t)$. Moreover, $J_{0}(x) = log\mathbb{E}[x] - \mathbb{E}[log(x)]$ denotes  the unconditional entropy measure (See Backus, Chernov, and Zin (2014),  Ghosh, Julliard, and Taylor (2011), Stutzer (1995) for applications of entropy to asset pricing).


\subsection{A long-run law of entropy}

For a non-negative process $(Y_t)$, assuming the existence of the following limit and we include $+\infty$ as plausible limit, define 
\be
z_{\infty}(Y) = \lim_{n \rightarrow \infty} \frac{ J_0(Y_1) + J_1(Y_1)  + \cdots + J_{n-1}(Y_n)}{n}, P-a.s.
\ee
$z_{\infty}(Y)$ is the {\em long-term entropy} of $(Y_t)$.

\begin{pr}
\label{pr:multiplicative}
For a general positive process $(Y_t)$ with Assumption I, 
\be
\limsup_{n \rightarrow \infty} V_n(Y) \le 1, P-a.s.
\ee
If Assumption II holds for the process $(log(Y_t))$, and $z_{\infty}(Y)$ exists, then
\be
\lim_{n \rightarrow \infty} V_n(Y) = e^{ -z_\infty(Y)}, P-a.s.
\ee
\end{pr}

%

The first part of Proposition \ref{pr:multiplicative} is non-trivial, and it essentially implies the classical Dybvig, Ingersoll and Ross (1996)' long forward rate theorem (see its proof in Appendix A). By the multiplicative Doob-Meyer decomposition as follows (Williams, 1991), 
\be
Y_n = L_n B_n, L_n = \prod_{t=1}^{n} \left(\frac{Y_t}{\mathbb{E}_{t-1}[Y_t] } \right), B_n = Y_0 \prod_{t=1}^{n} \left( \frac{ \mathbb{E}_{t-1}[Y_t]}{Y_{t-1}} \right)
\ee
where $(L_n)$ is a martingale and $(B_n)$ is predictable process. Therefore, the long-term growth rate of a general process ($Y_t$) is bounded  above by the long-term growth rate of its predictable component, that is, 
\be
\limsup_{n \rightarrow \infty} \left(Y_{n}\right)^{\frac{1}{n}} \le \limsup_{n \rightarrow \infty} \left(B_{n}\right)^{\frac{1}{n}}.
\ee

More importantly, the second part of Proposition \ref{pr:multiplicative} determines the long-term growth rate of a general positive martingale precisely. If the conditional variance of $log(Y_t)$ have a bounded expectation, or alternatively, the conditional variance between $log(Y_t)$ and its one-step ahead forecasting $\mathbb{E}_{t-1}[log(Y_t)]$ is bounded, then the long-term growth rate of a martingale is its long-term entropy. 

 
%

\begin{pr}
	\label{pr:existence}
	Assuming $(Y_t = X_1 \cdots X_{t})$ is a positive multiplicative martingale, that is, $\mathbb{E}_{t-1}[X_t] = 1$ and $X_t$ is ${\cal F}_t$-adapted. If $( log(Y_t) )$ satisfies Assumption II, then, there exists a subsequence $t_1< t_2  < \cdots $ and a positive random random $\zeta$ such that
	\be
	\lim_{n \rightarrow \infty} \left( X_{t_1} \cdots X_{t_n}\right)^{\frac{1}{n}} = \zeta, a.s.
	\ee
\end{pr}

Proposition \ref{pr:existence} shows the existence of the long-term growth rate in a weaker sense. Imposing further technical assumption, Proposition \ref{pr:existence} implies the existence of long-term entropy. Therefore, in the subsequent discussions, we do not document these technical conditions but simply assume the existence of the long-term entropy.

\begin{exa}
Let $Y_t = exp(a + \sigma \zeta_t), \sigma > 0$, and IID, $\zeta_t \sim {\cal N}(0, 1)$. Then 
$z_{\infty}(Y) = \frac{1}{2} \sigma^2$. On the other hand, for a martingale $(M_t)$ in Alvarez and Jermann (2005, Example 4.3), $x_{t+1} = log M_{t+1} = log \beta + \rho log M_t + \epsilon_{t+1}$, $0 < \rho < 1$, and IID, $\epsilon_t \sim {\cal N}(0, 1)$. Then
$(M_{t})^{\frac{1}{t}} \rightarrow 1, a.s.$
\end{exa}

\subsection{Implication to long-dated pricing}

We consider a risk-adjusted asset return process $( M_t R_{0,t})$ for a return process $(R_t)$ and a pricing kernel process $(M_t)$. In the following discussions, the asset's gross return is always strictly positive, so $log (R_{t})$ and the relevant long-term entropy is well defined.


\begin{cor}
\label{cor:long-dated}
If there exists a positive number $L$ such that $\mathbb{E}\left[Var_{t-1}(log (m_t R_t)) \right] \le L, \forall t$, and $z_{\infty}(mR)$ exists, then 
\be
\lim_{n \rightarrow \infty} \left( M_n R_{0, n}  \right)^{\frac{1}{n}} = e^{-z_{\infty}(mR)}.
\ee
In particular, if $\mathbb{E}[Var_{t-1}(log (m_t))] \le L, \forall t$, and $z_{\infty}(m)$ exists, then for $R_{0, n-1}^{f} = R_{f,0} \cdots R_{f, n-1}$,
\be
\lim_{n \rightarrow \infty} \left( M_n R_{0, n-1}^{f}  \right)^{\frac{1}{n}} = e^{-z_{\infty}(m)}
\ee
Moreover, if the long-term growth rate of the pricing kernel process $ \lim_{n \rightarrow \infty} \frac{log (M_n)}{n}$ exists, then there exists a limit, 
\be
\label{eq:pricing-kernel-multiplicative}
r_{f, \infty}  = \lim_{n \rightarrow \infty} \frac{ \sum_{t=1}^{n} r_{f, t-1}}{n} = -z_{\infty}(m) - \lim_{n \rightarrow \infty} \frac{log(M_n)}{n}, a.s.
\ee
\end{cor}

As stated in Section 3.3, in a non-generic case, 
 the sequence $m_n R_n \rightarrow 1, a.s.$, thus
$\left( m_1 R_1 \cdots m_n R_n \right)^{\frac{1}{n}} \rightarrow 1$. 
 In this case, the asset is asymptotically optimal growth portfolio and the pricing kernel is the reciprocal of the optimal growth portfolio. However, for the generic case, as shown in Martin (2012), $\left( m_1 R_1 \cdots m_n R_n \right) \rightarrow 0, a.s.$.  Corollary \ref{cor:long-dated} is stronger in that the geometrical average, $V_n(mR)$, converges to $e^{-z_{\infty}(mR)}$, almost surely.

%

 Again, if the short rate process $(r_{f,t})$ is stationary and ergodic, then $\overline{r}_{\infty} = \lim_{n} r_{f,n}$ exists and $r_{f, \infty} = \overline{r}_{\infty}$. Interestingly, under certain condition about the stochastic discount factor, we obtain the existence of $r_{f, \infty}$. For simplicity, $r_{f, \infty}$ is called a {\em long-term short rate}.

 

\subsection{Characterization of $z_{\infty}(m)$}

The next result characterizes the long-term entropy of the pricing kernel using the excess asset return.

\begin{cor}
\label{cor:entropy}
In a no-arbitrage financial market with a stochastic discount factor process $(m_t)$, for any asset return $R_t$ such that $\mathbb{E}[Var_{t-1}(log R_t)]$ is uniformly bounded above by a positive constant, 
\beq
z_{\infty}(m) \ge \lim_{n \rightarrow \infty} \frac{ \sum_{t=1}^{n} (r_t - r_{f,t-1})}{n}.
\eeq
Moreover, if there exists a positive number $L$ such that $\mathbb{E}\left[Var_{t-1}(\frac{1}{m_t})\right] \le L, \forall t$, then
\be
z_{\infty}(m) = \sup_{R}  \lim_{n \rightarrow \infty} \frac{ \sum_{t=1}^{n} (r_t - r_{f,t-1})}{n},
\ee
where $R$ runs through processes of asset returns, that is, $(m_1R_1 \cdots m_t R_t$) is a martingale.
\end{cor}

This result states that the long-term entropy of the stochastic discount factor must be bounded from below by any risky asset's long-term excess return (in continuously compounding). It is remarkable to compare this characterization of the long-term entropy with the duality theorem in Hansen-Jaganathan (1991) on the variance of the stochastic discount factor and Cochrane and Saa-Requejo (2000)'s good-deal bound on the stochastic discount factor.
There are several significant points in Corollary \ref{cor:entropy}. First, $z_{\infty}(m)$ is about the dynamic (long-term) property of the pricing kernel whereas previous studies focus on the one-period stochastic discount factor. Second,  the long-run excess mean in Corollary \ref{cor:entropy} is computed by the realization data. Then, regardless of any no-arbitrage asset pricing models, the long-term entropy equals the maximum long-run excess mean in continuously compounding. For a long-lived agent, the long-term entropy can be calculated by sample data only. By contrast, the future distribution assumption is required in Hansen-Jaganathan (1991), Stutzer (1995), Cochrane and Saa-Requejo (2000), Almeida and Garcia (2017), and Liu (2021). Third, from a long-lived agent's perspective, the long-term entropy in any no-arbitrage asset pricing model (under assumptions in Corolllary \ref{cor:entropy}) should be fairly close to the long-term excess return in continuously compounding. 

\begin{figure}
\begin{center}
\caption{A comparison between a disaster and a long run risk model}
\label{fig:moments}
\rule{0pt}{3pt}
\parbox{7in}
\medskip
\parbox{6.5in}
{\small 
This figure displays the long-term higher moments of stochastic discount factor in Corollary \ref{cor:liu-dual} for negative (resp. positive) value of $s$ in top panel (bottom panel), in a disaster model and a long run risk model.
For a disaster model. I use the values of parameters calibrated in Liu (2021), $\beta = 0.99, \gamma = 4, r_f = 0.02, \sigma = 0.02, \nu = 0.23, \omega = 0.017, \theta = -0.38$. The results are essentially the same if I use other mild or severe choices of $(\omega, \theta)$. In the long run risk model, I use the parameters calibrated in Bansal and Yaron (2004), Bansal, Kiku and Yaron (2012): $\mu = 0.0015, \rho = 0.979, \sigma = 0.0078, \phi_{e} = 0.44, \nu_1 = 0.987, \delta = 0.998, \sigma_{w} = 0.23 * 10^{-5}$ and $\gamma = 10$. By calculation, the parameter $\theta = \frac{1-\gamma}{1-\frac{1}{\psi}} = 4, \kappa_0 =  3.266$ and $\kappa_1 = 0.997$. Since I only consider the pricing kernel process, I do not need the process of $g_{d, t+1}$. This figure shows that the long run risk model performs better than the disaster model for both negative and positive value of the parameter $s$ in the line by Corollary \ref{cor:liu-dual}.}
\bigskip
\includegraphics[width=6in, height=2.5in]{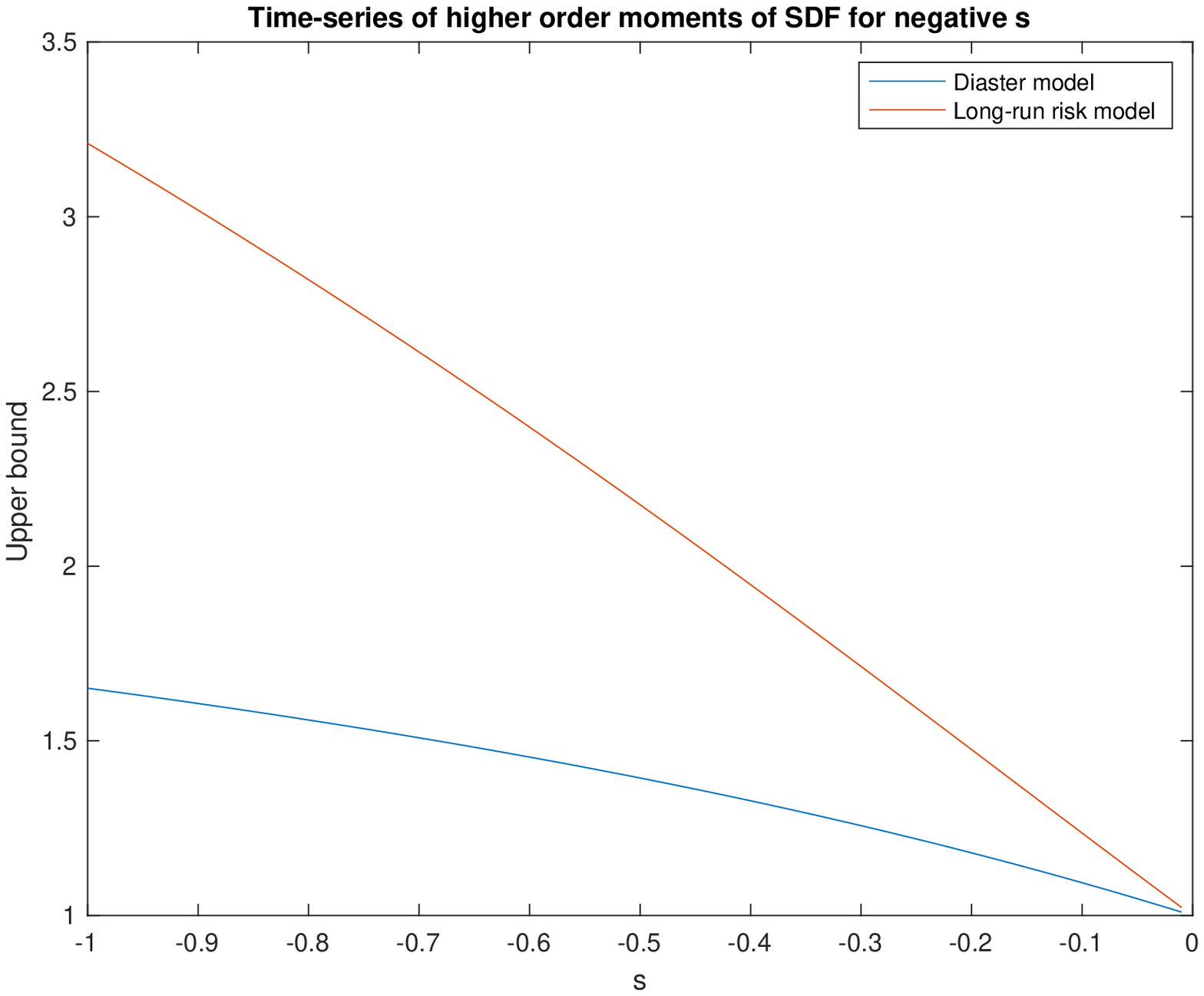}
\includegraphics[width=6in, height=2.5in]{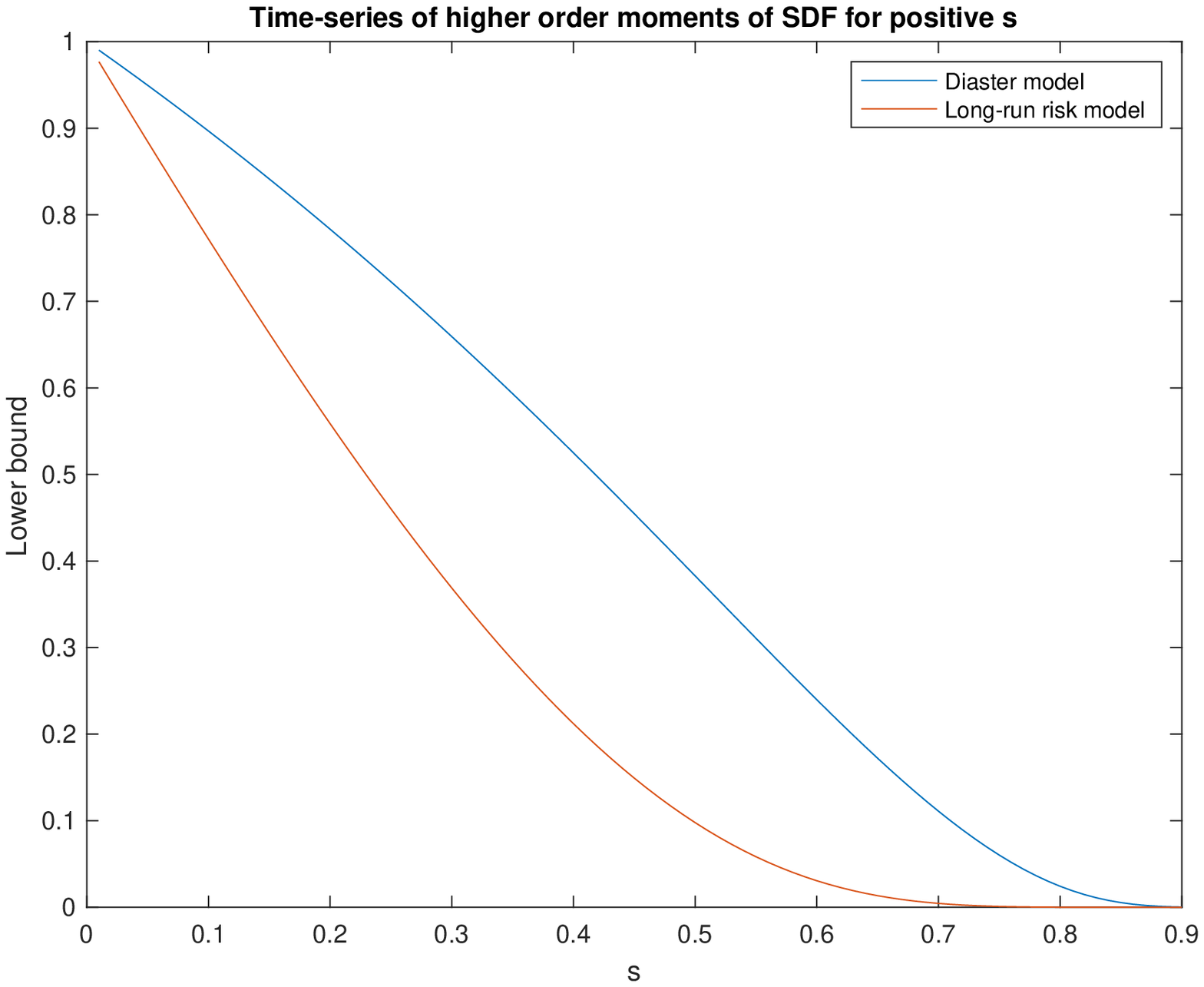}
\end{center}
\end{figure}

\begin{figure}
\begin{center}
\caption{The long-term entropy between a disaster and a long run risk model}
\label{fig:entropy}
\rule{0pt}{3pt}
\parbox{7in}
\medskip
\parbox{6.5in}
{\small 
This figure displays the long-term entropy of the pricing kernels in a disaster model and a long run asset pricing model. I use the same specification and model parameters of the pricing kernels  as in Figure \ref{fig:moments}. As shown, the long-term entropy $z_{\infty}(m^s)$ in a long run model is smaller than that in a disaster model. Indeed, $z_{\infty}(m) = 0.015 $ in the long run risk model. By Corollary \ref{cor:entropy}, the long-run excess mean of monthly asset return (continuously compounding)  is bounded by 1.5\%, which is clearly too small. In the disaster model, the long-term  entropy is $z_{\infty}(m) = 0.0885$. Equivalently, the long-run excess mean of monthly asset return (continuously compounding) is bounded by 8.85\%. The reason of a small long-term entropy in the the long-risk model is due to a too high long-term short rate $\overline{r}_{\infty}^{f} = 2.3166$ (annually). In contrast, the long-term short rate is 2 \%. Therefore, a better asset pricing model should have a small long-term short rate but a large long-run excess mean of asset return, from a long run perspective.
}
\bigskip
\bigskip
\includegraphics[width=6in, height=2.5in]{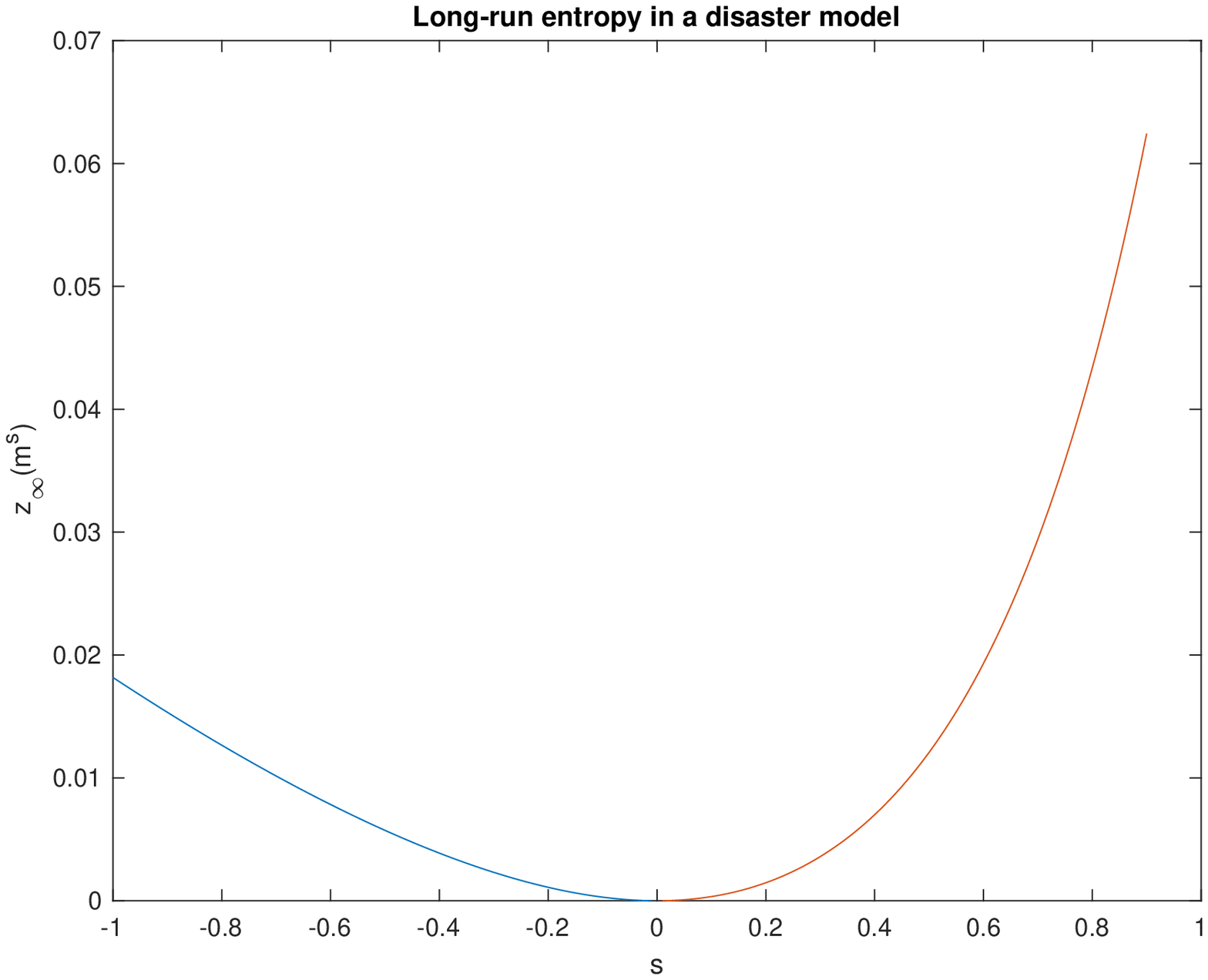}
\includegraphics[width=6in, height=2.5in]{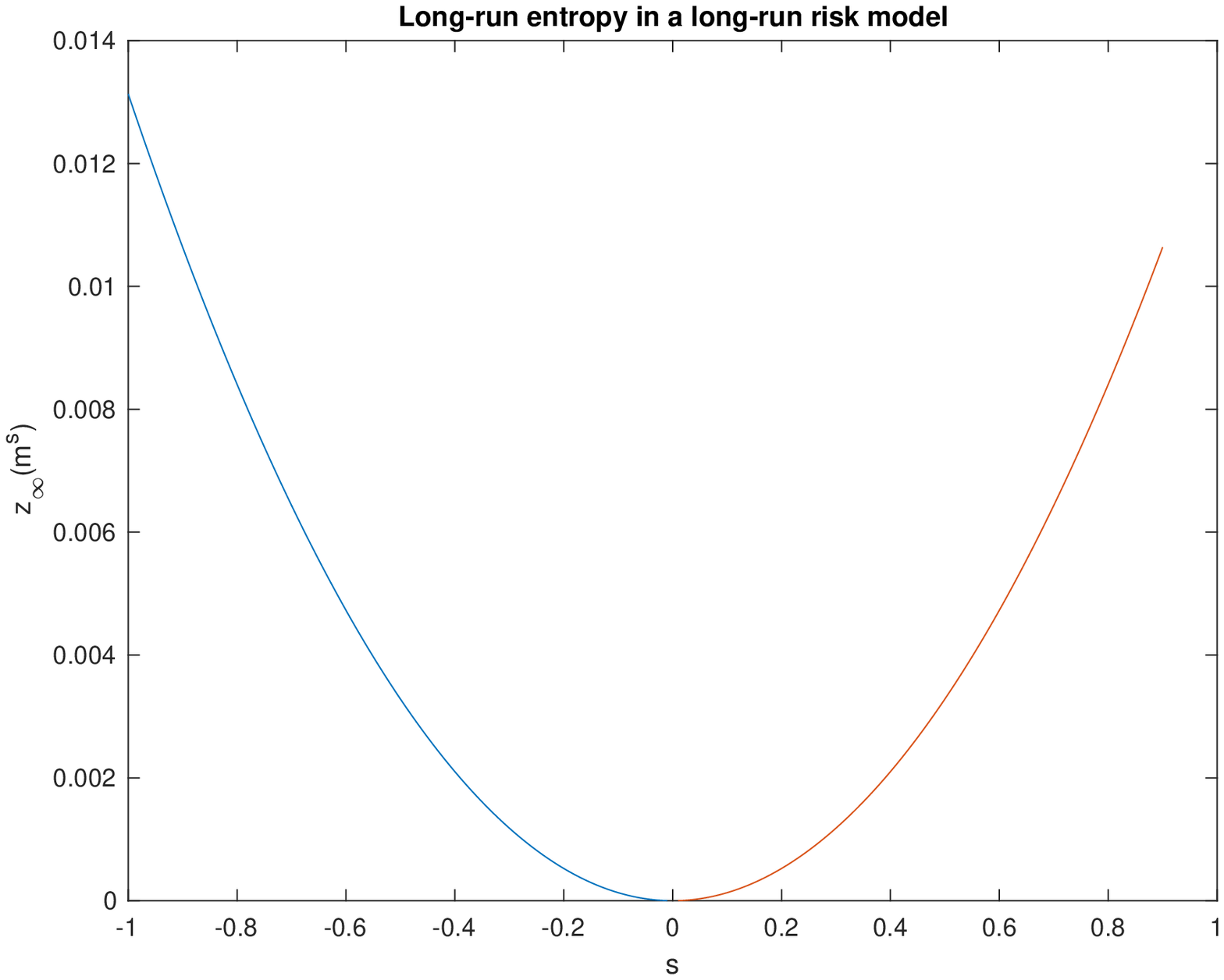}
\end{center}

\end{figure}

Following the discussions in Section 3.2, we use Corollary \ref{cor:entropy} to compare disaster and long risk model again. Figure \ref{fig:entropy} displays $z_{\infty}(m^s)$ for a disaster model and a long run risk model. In both disaster model and long run risk model, $z_{\infty}(m)$ is calculated by $\mathbb{E}_{t-1}[log m_t]$ as in Example 4.1 and Example 4.2. we apply the same parameters as in Figure \ref{fig:moments}. By Corollary \ref{cor:entropy}, a larger value of the long-term entropy in an asset pricing model is better to fit the long-run excess mean of assets (in continuously compounding). Remarkably, this comparison yields a different message from that in Section 3 (Figure \ref{fig:moments}). The long-term entropy in a long risk model is 0.015, showing that the long-run excess mean of monthly asset return (continuously compounding) is bounded by 1.5\%. By contrast, for the calibrated disaster model, the long-term entropy is 0.0885, showing the long-run excess mean of monthly asset return is bounded by 8.85\% (in continuously compounding). The disaster model is better than the calibrated long run model by investigating the long-run excess mean of monthly asset return in the stock market. 

Why is the long run risk model performs not as good as a disaster model  with this measure? One possible reason is that the persistence of the state process in the long run risk model is not high enough to yield a large value of the long-term entropy. In a long run risk model, a highly persistent state price plus an investor's early resolution of risk affect long run model outcomes. To obtain a significant value of the long-term entropy, the state process must be highly persistent such as stochastic volatility (Pohl, Schmedders, and Wilms (2018)). On the other hand, the jump component in a disaster model ensures the highly persistence of the state price, thus a large value of the long-term entropy. The different implications in Figure \ref{fig:moments} and Figure \ref{fig:entropy} show that these two long-term measures, $z_{\infty}(m^s)$ and $\pi(m;s)$, play {\em critical yet different role} in diagnosing asset pricing models. It suggests the importance of significant persistent component for more complicated state process or Bayesian learning about the key parameters (both the expected return and variance)  such as in Collin-Dufresne, Johannes, and Lochstoer (2016, 2017), Weitzman (2007).

\subsection{Alternative long-term measures}

In this subsection, we discuss the relation between the long-term entropy and other long-term measures in earlier literature (Hansen (2012), and Backus, Chernov, and Zin (2014)). 

Following Backus, Chernov, and Zin (2014), define 
\be
I_{t}(n) = \frac{1}{n} \mathbb{E}[J_{t}(m_{t,t+n})].
\ee
In particular, for $n = 1$, since $log \mathbb{E}_{t}[m_{t+1}] = -r_{f,t}$, the {\em shortest-horizon entropy} is 
\beq
I_{t}(1) = - \mathbb{E}[log (m_{t+1})] - \mathbb{E}[r_{f, t}].
\eeq
 Moreover, if the limit exists, 
\be
I_{t}(\infty) = \lim_{n \rightarrow \infty} I_{t}(n) = \lim_{n \rightarrow \infty} \frac{\mathbb{E}[ J_{t}(m_{t, t+n})]}{n}.
\ee
Different from Backus, Chernov, and Zin (2014) in notations, we use the script ``t" to represent the conditional on time $t$ to calculate the entropy before computing its unconditional mean. 
%
Define the long-term yield at time $t$ by
\be
y_{t}^{\infty} = \lim_{n \rightarrow \infty} y_{t}^{t+n}.
\ee
Notice that $y_{t}^{ \infty} = - log \left( \lim_{T \rightarrow \infty} D(t, T)^{\frac{1}{T-t}} \right) = log(1 + z_{L}(t))$, where 
\beq
z_{L}(t) \equiv \lim_{T \rightarrow \infty} \{ D(t, T)^{-\frac{1}{T-t}} - 1\}
\eeq
 is the {\em long zero-coupon rate} introduced in  Dybvig, Ingersoll and Ross (1996). 
 Lastly, we define the long-term growth (or decay) rate
\be
\rho_{t}(M) = \lim_{n \rightarrow \infty} \frac{log \mathbb{E}_{t}[M_{n}/M_{t}]}{n},
\ee
which is an conditional version of the long-term rate in Hansen (2012). Since $ \mathbb{E}_{t}[M_{n}/M_{t}] = D(t, n)$, it is easy to see that
\beq
\rho_{t}(M) =  \lim_{n \rightarrow \infty} \frac{log D(t, n+t)}{n} = -y_{t}^{\infty}.
\eeq

\begin{lem}
\label{lem:BCZ}
Assuming that $(\mathbb{E}_{t}[log (m_{t+1})])$ and $(r_{f,t})$ are stationary and ergodic, then $z_{\infty}(m) = I_{t}(1), \forall t$. 
In general, under regularly conditions, 
\be
\label{eq:BCZ}
I_{t}(\infty) = \mathbb{E}[z_{\infty}(m)] + \mathbb{E}\left[ r_{f, \infty} - y_{t}^{\infty} \right],
\ee
Moreover, $I_{t}(\infty)$ and $\rho_{t}(M)$ are  {\em decreasing} with respect to $t$.
\end{lem}

Lemma \ref{lem:BCZ} is useful for long-run analysis. It states that $z_{\infty}(m)$ equals to $I(1)$ in those asset pricing models in a stationary and ergodic environment. According to Corollary \ref{cor:entropy}, a large value of the long-term entropy is appealing to bound the long-run excess return of any asset. It is consistent with Backus, Chernov, and Zin (2014) to demand a large value of $I(1)$. 

Lemma \ref{lem:BCZ} also documents how the measure $I_{t}(\infty)$ depends on the time variable $t$ which is not discussed in Backus, Chernov and Zin (2014). Since the long forward rate never fall in an arbitrage-free market (Dybvig, Ingersoll and Ross, 1996), the sequence $y_{t}^{\infty}$ never fall with time. Therefore, $I_{t}(\infty)$ is non-increasing with time $t$. Equation (\ref{eq:BCZ}) presents an important relationship among four long-term measures,  $\left\{
I_{t}(\infty), z_{\infty}(m), r_{f, \infty}, y_{t}^{\infty} \right\}.$



\begin{exa}
Assuming the interest rate is constant, then by Lemma \ref{lem:BCZ}, $I(\infty)  = z_{\infty}(m)$. In particular, for Campbell and Cochrane (1991)'s internal (difference) habit model with constant interest rate (with their particular choice of $\lambda(s_t)$), $z_{\infty}(m) = I(\infty)$. On the other hand, for the Chan and Kogan (2002)'s internal (ratio) habit model with a representative agent, the equilibrium interest rate is a constant (Chan and Kogan (2002, Lemma 7)). Then in this representative Chan and Kogan's model, $z_{\infty}(m) = I(\infty)$. 
\end{exa}

In Campbell and Cochrane (1991)'s internal (difference) habit mode, $z_{\infty}(m) = I(\infty) = 0.023$,  as shown in Backus, Chernov and Zin  (2014, Table III). By using Corollary \ref{cor:entropy}, it states that the long-run excess monthly return is bounded above by 2.3 percent (and annually 27.6 percent), which is reasonable in a long run given the market's excess return is about 8 percent annually. Both the habit model and the disaster model offer reasonable long-term entropy to fit the market data. 

In contrast, in Chan and Kogan's representative agent ratio habit model, the long-term entropy seems too small, around 0.03 percent per month and 0.36 percent annually (See  Backus, Chernov and Zin (2014, Table II)). It means that the equilibrium ratio model with a representative agent version is not able to capture the long-term sample excess return of assets in the equity market. Moreover, Xiouros and Zapatero (2010) document that the heterogeneous equilibrium model in Chan and Kogan (2002) is also unlikely be able to explain several empirical regularities, and suggest the importance of a varying conditional volatility of the state variable as in the internal (difference) habit model.

\subsection{Permanent and Temporary  of SDF}


Following Alvarez and Jerman (2005), define
\beq
R^{f}_{t+1, k} = \frac{D(t+1, t+k)}{D(t, t+k)}
\eeq
as one-period holding return at time $t$ on a zero-coupon bond maturing $t+k$. Then, the limit when $k \rightarrow \infty$ is denoted by $R_{t+1, \infty}^{f}$, the one-period holding return on a bond with infinite maturity.
Then $\mathbb{E}_{t}[m_{t+1} R^{f}_{t+1,k}] = 1, \forall k \ge 1$. 
 Under regularity conditions (Assumption 1 and 2 in Alvarez and Jerman (2005), or the existence of dominated positive eigenvalue in Hansen and Scheinkman (2009), Hansen (2012)), there exists 
  $R^{f}_{t+1, \infty} = \lim_{k \rightarrow \infty} R^{f}_{t+1, k}, a.s,$ 
 as the one-period return on a bond with infinite maturity. Moreover, $\mathbb{E}_{t}[m_{t+1} R^{f}_{t+1, \infty}] = 1$.\footnote{Under regular condition, Lebesgue's dominance theorem implies that $\mathbb{E}_{t}[m_{t+1} R^{f}_{t+1, \infty}] = \lim_{k \rightarrow \infty} \mathbb{E}_{t}\left[m_{t+1} R^{f}_{t+1,k}\right] = 1$.}
  Therefore,  there exists a multiplication decomposition $
m_{t+1} = m_{t+1}^{P} m_{t+1}^{T}$, 
where $m_{t+1}^{P} = m_{t+1} R_{t+1, \infty}^{f}$ measures the {\em permanent  component} of the stochastic discount factor, and $ m_{t+1}^{T} = \frac{1}{R_{t+1, \infty}^{f}}$ is the {\em temporary component.}
Given a pricing kernel process $(M_t)$, its permanent component is $
M_{n}^{P} \equiv M_{n} \prod_{t=1}^{n} R_{t,\infty}^{f}$. 
The permanent component $(M_n^P)$ is a martingale. Notice that $R^{f}_{t+1, k}$ is only known at time $t+1$ for all $k \ge 2$,  then $R_{t, \infty}^{f}$ is ${\cal F}_{t+1}$-adapted whereas $R_{f, t}$ is ${\cal F}_t$-adapted. This is a crucial difference between the decomposition in Alvarez and Jerman (2005) and Hansen and Scheinkman (2009), and the multiplicative Doob-Meyer decomposition. That is, the temporary component is {\bf not} predictable since it involves the bond market information at {\em future} time.

\begin{cor}
\label{cor:AJ}
Assume both processes $( log(M_n))$ and $(log(M_n^P))$ satisfy Assumption II, that is, $
\mathbb{E}[Var_{t-1}(log (M_t))], \mathbb{E}[Var_{t-1}(log( M_t^P))] < L,\forall t.$ Moreover, $z_{\infty}(M)$, $r_{f, \infty}$ and the limit $\delta = \lim_{n \rightarrow \infty} \frac{ log R_{1, \infty}^{f} + \cdots + log R_{n, \infty}^{f}}{n}, a.s.,$ exist, then 
\be
\label{eq:AJ}
\lim_{n\rightarrow \infty} (M_{n}^{P}) ^{\frac{1}{n}} = e^{-z_{\infty}(M) + \delta - r_{f, \infty}}.
\ee
Under Assumption II for asset return process $( log R_t )$ and $( log R_{t, \infty}^{f})$, then
\be
\label{eq:AJ2}
z_{\infty}(M^{P}) \ge log \lim_{n \rightarrow \infty} \left( \frac{R_1 \cdots R_{n}}{R_{1, \infty}^{f} \cdots R_{n, \infty}^{f}} \right)^\frac{1}{n}.
\ee
Moreover, if there exists a positive number $L$ such that $\mathbb{E}\left[Var_{t-1}(\frac{1}{m_t})\right] \le L$, then
\be
\label{eq:AJ3}
z_{\infty}(M^{P}) = \sup_{R} log \lim_{n \rightarrow \infty} \left( \frac{R_1 \cdots R_{n}}{R_{1, \infty}^{f} \cdots R_{n, \infty}^{f}} \right)^\frac{1}{n} = \sup_{R} \lim_{n \rightarrow \infty} \frac{ \sum_{t=1}^{n} (r_{t} - r^{f}_{t, \infty})}{n},
\ee
where $R$ runs through all asset returns process such that $M_t R_{0, t}$ is a martingale, and $r^{f}_{t, \infty} = log R^{f}_{t, \infty}$.
\end{cor}

According to Corollary \ref{cor:AJ},  the long-term entropy of the permanent pricing kernel is 
\be
\label{eq:z-AJ}
z_{\infty}(M^P) = z_{\infty}(M) - \delta + r_{f, \infty}.
\ee


Empirically speaking,  the size of the number $\delta - r_{f, \infty}$  is very small from the bond market, so $z_{\infty}(M^P)$ is very close to the long-term entropy. Indeed, Equation (\ref{eq:AJ3}) states that the long-term entropy is the maximum long-run excess asset return over the infinite-maturity (console) bond return. Finally, by its definition, the number $\delta$ is the sample average of the continuous return, $R_{t, \infty}^{f}$.  Compared with the long-term short rate $r_{f,\infty}$, the number $\delta$ concerns the return of long-term bond. 

\section{Conclusion}
\label{sec:conclusion}

This paper develops the additive and multiplicative version of the long-run law of unexpected shocks for economic variables. These long-run laws of unexpected shocks rely upon only a uniform upper bound of the unconditional variance of the shocks, and this condition is also necessary to derive meaningful asymptotic results. The asset pricing implications of the long-run laws are related to some essential insights of the following theories. (1) The long-dated asset valuation and tail event analysis in the long-term (Martin, Weitzman, Nordhaus). (2) The long-run theory of stochastic discount factor and risk-adjusted asset return (Hansen and Scheinkman). (3) The measures of the stochastic discount factors (Hansen and Jagannathan, Alvarez and Jermann). (4) The no-arbitrage asset pricing theory and the long-run forward rate (Dybvig, Ingersoll, and Ross), and (5) The comparison of subjective, objective, and risk-neutral probability (Nagel).

The long-run analysis implies several long-term measures such as $\pi(m, s)$ and $z_{\infty}(m)$. We characterize these measures in terms of sample data of asset returns and interest rate only. Moreover, we use these new characterizations to several leading asset pricing models. These results suggest the importance of these long-run laws to non-ergodic and non-stationary economies.


%

\newpage

\renewcommand {\theequation}{A-\arabic{equation}} \setcounter
{equation}{0}
\section*{Appendix A. Proofs  of Propositions}	

In this Appendix, I present the proofs of major results. In the Online Appendix I provide the proofs of other propositions and all corollaries.
%

{\bf Proof of Proposition \ref{pr:additive} and Proposition \ref{pr:additive-weighted}.}


{\em Claim: For any monotonic positive real numbers sequence $b_n \uparrow +\infty$, $c_n \downarrow 0$ with $\sum_{n=1}^{\infty} c_n^2 < \infty$, we have
\be
\label{eq:extension}
\frac{1}{b_n} \sum_{k=1}^{n} (b_k c_k) (Y_k - \mathbb{E}_{k-1}[Y_k]) \rightarrow 0, a,s.
\ee
Moreover, it converges to 0 in $L^2(\Omega)$ if each $Y_n \in L^2(\Omega)$.}

On the one hand, choosing $b_n = n^{\alpha}, c_n = \frac{1}{n^{\alpha}}, \alpha > \frac{1}{2}$, we obtain Proposition \ref{pr:additive}. 
On the other hand, let $c_k  = \frac{1}{k}, b_k = k a^{-k}$ then $\sum_{n} c_n^2  < \infty$ and $b_n \uparrow \infty$ since $0 < a \le 1$. Then, 
\beq
 \frac{1}{n a^{-n}} \sum_{k=1}^{n} a^{-k} (Y_k - \mathbb{E}_{k-1}[Y_k])  =  \frac{1}{n} \sum_{k=1}^{n} a^{n-k} (Y_k - \mathbb{E}_{k-1}[Y_k])  \rightarrow 0, a.s.
\eeq
This leads to Proposition \ref{pr:additive-weighted}.


It remains to prove the ``Claim".  Define $Z_k = Y_k - \mathbb{E}_{k-1}[Y_k], \mathbb{E}_{k-1}[Z_k] = 0$, and $
\tilde{U}_n = \sum_{k=1}^{n} c_k (Y_k - \overline{Y}_k), n = 1, \cdots.$
By its definition, $\tilde{U}_n$ is a martingale. Moreover,
\begin{eqnarray}
\tilde{U}_{n}^2 & = & \sum_{k=1}^{n} c_k^2  Z_k^2 + 2 \sum_{i < j} c_i c_j Z_i Z_j.
\end{eqnarray}
By Assumption II that $\mathbb{E}[Z_k^2] = \mathbb{E}[ Var_{k-1}(Y_k)] \le L, \forall k$ and the iterated law of the conditional expectation, we have $\mathbb{E}[Z_i Z_j] = \mathbb{E}[Z_i \mathbb{E}_{i+1}[Z_j]] = 0$, and thus
\be
\mathbb{E}[\tilde{U}_n^2] \le \sum_{k=1}^{n} c_k^2  L \le L \sum_{k=1}^{\infty} c_k^2.
\ee
Then, by the Doob's martingale convergence theorem (William, 1991), $\tilde{U}_n$ converges almost surely to a finite variable with finite moment. By the Kronecker lemma (William, 1991), the sequence $\frac{1}{b_n} \sum_{k=1}^{n} (b_k c_k) Z_k$ converges to zero almost surely. If each $Y_n \in L^2(\Omega)$, then $\tilde{U}_n \in L^2(\Omega)$, Then by the Doob's martingale convergence theorem again, the sequence $\frac{1}{b_n} \sum_{k=1}^{n} (b_k c_k) Z_k$ converges to zero in $L^2(\Omega)$. Then our result follows from a $L^2$-type Kronecker lemma, which proof can be easily modified from Willams (1991). \hfill$\Box$

%

To prove the first part of Proposition \ref{pr:multiplicative} in a general situation, we need the following lemma, which belongs to Hubalek, Klein, and Teichmann (2002).
\begin{lem}
	\label{lem:HKT}
	Given a non-negative random variable sequences $X_n $ and $X_n \rightarrow X_{\infty}, a,s.$.  If $\liminf_{n \rightarrow \infty} \mathbb{E}[X_{n}^{n}]^{\frac{1}{n} } = C < \infty$, a.s., then $X \le C, a.s.$.
\end{lem}

{\bf Proof of Proposition \ref{pr:multiplicative}.}

Let $X_n = \left(m_1 R_1 \cdots m_n R_n \right)^{\frac{1}{n}}$, then $X_n^{n} = m_1 R_1 \cdots m_n R_n$, so $C = \liminf_{n \rightarrow \infty} \mathbb{E}[X_{n}^{n}]^{\frac{1}{n} }= 1$. Then, any convergence subsequence of  $(X_n)$ has a limit $X\le 1$. It implies that $\limsup_{n} X_n \le 1$. 

For the first part, let $X_n = D_n = \left( \prod_{t=1}^{n} \frac{ Y_t}{\mathbb{E}_{t-1}[Y_t]} \right)^{\frac{1}{n}}$, then $D_{n}^n = \prod_{t=1}^{n} \frac{ Y_t}{\mathbb{E}_{t-1}[Y_t]}$. Then, by the iterate law of expectation, we obtain
\beq
\mathbb{E}\left[ D_n^n\right] = \mathbb{E}\left[ \frac{ Y_t}{\mathbb{E}_{t-1}[Y_t]} \right] = 1.
\eeq
Therefore, by Lemma \ref{lem:HKT}, we have shown that $\limsup_{n \rightarrow \infty} D_n \le 1$. The second part follows from Proposition \ref{pr:additive} and the definition of $z_{\infty}(Y)$. \hfill$\Box$

{\bf Proof of Proposition \ref{pr:existence}.}

The Doob-Meyer decomposition of $log(Y_t)$ is $M_t + A_t$, where $M_t = S_1 + \cdots + S_t, S_u = log(Y_u) - \mathbb{E}_{u-1}[log(Y_u)];  A_t = T_1 +\cdots +T_t, T_u =  \mathbb{E}_{u-1}[log(Y_u)] - log(Y_{u-1})$. By using the Weizsacker-Kolmos' theorem (Weizsacker, 2004) for non-positive random variables $T_u$, there exists a subsequtence $t_1 < t_2 < \cdots $ such that $\frac{T_{t_1} + \cdots + T_{t_n}}{n} \rightarrow \zeta, a.s.$. Moreover, by the same proof of Proposition \ref{pr:additive}, we can show that $\frac{S_{t_1} + \cdots + S_{t_n}}{n} \rightarrow 0, a.s.$ (under Assumption II). Therefore, $\frac{S_{t_1} + T_{t_1} + \cdots + S_{t_n} + T_{t_n}}{n} \rightarrow 0$. Notice that $S_u + T_u = log(Y_u)  - log(Y_{u-1}) = log(X_u)$. The proof is finished.\hfill$\Box$

\newpage

\newpage


\newpage

\newpage

%
%

%
%
%
%
%

\end{document}